\documentclass[aps,superscriptaddress,reprint,nofootinbib]{revtex4-1}

\usepackage{graphicx}
\usepackage{amsmath}
\usepackage{amssymb}
\usepackage{amsfonts}
\usepackage{filecontents}
\usepackage{color}
\usepackage[normalem]{ulem} 
\usepackage{soul}
\usepackage{epstopdf}
\usepackage{nameref}
\usepackage{hyperref}
\hypersetup{hidelinks}
\usepackage{fancyhdr}
\usepackage{float}
\usepackage{gensymb}
\usepackage{physics}
\usepackage{upgreek}
\newcounter{fig}
\usepackage{lipsum}
\usepackage{natbib}
\usepackage{bibunits}

\usepackage[usenames,dvipsnames,svgnames,table]{xcolor}
\usepackage{enumitem}
\usepackage{multirow}

\DeclareRobustCommand{\SkipTocEntry}[5]{}

\defaultbibliography{Raman_paper_reference.bib}
\defaultbibliographystyle{naturemag}

\renewcommand{\bibliography}[1]{} 

\begin{document}
\begin{bibunit}[naturemag]

\title{All-optical ultrafast arbitrary rotation of hole orbital qubits with direct phase control}

\author{Jun-Yong Yan}
\affiliation{State Key Laboratory of Extreme Photonics and Instrumentation, College of Information Science and Electronic Engineering, Zhejiang University, Hangzhou 310027, China}

\author{Liang Zhai}
\affiliation{Institute of Fundamental and Frontier Sciences, University of Electronic Science and Technology of China, Chengdu 610054, China}
\affiliation{Department of Physics, University of Basel, Basel CH-4056, Switzerland}

\author{Hans-Georg Babin}
\affiliation{Lehrstuhl für Angewandte Festkörperphysik, Ruhr-Universität Bochum, Bochum DE-44780, Germany}

\author{Yuanzhen Li}
\affiliation{State Key Laboratory of Extreme Photonics and Instrumentation, College of Information Science and Electronic Engineering, Zhejiang University, Hangzhou 310027, China}
\affiliation{International Joint Innovation Center, Zhejiang University, Haining 314400, China}

\author{Si-Hui Pei}
\affiliation{College of Optical Science and
Engineering, Zhejiang University, Hangzhou 310027, China}

\author{\\Moritz Cygorek}
\affiliation{Condensed Matter Theory, Department of Physics, TU Dortmund, 44227 Dortmund, Germany}

\author{Wei Fang}
\affiliation{College of Optical Science and
Engineering, Zhejiang University, Hangzhou 310027, China}

\author{Fei Gao}
\affiliation{State Key Laboratory of Extreme Photonics and Instrumentation, College of Information Science and Electronic Engineering, Zhejiang University, Hangzhou 310027, China}
\affiliation{International Joint Innovation Center, Zhejiang University, Haining 314400, China}

\author{Andreas D. Wieck}
\author{\\Arne Ludwig}
\affiliation{Lehrstuhl für Angewandte Festkörperphysik, Ruhr-Universität Bochum, Bochum DE-44780, Germany}

\author{Chao-Yuan Jin}
\affiliation{State Key Laboratory of Extreme Photonics and Instrumentation, College of Information Science and Electronic Engineering, Zhejiang University, Hangzhou 310027, China}
\affiliation{International Joint Innovation Center, Zhejiang University, Haining 314400, China}

\author{Da-Wei Wang}
\affiliation{Zhejiang Province Key Laboratory of Quantum Technology and Device, School of Physics, Zhejiang University, Hangzhou 310027, China}

\author{Feng Liu}
\email[Email to: ]{feng\_liu@zju.edu.cn}
\affiliation{State Key Laboratory of Extreme Photonics and Instrumentation, College of Information Science and Electronic Engineering, Zhejiang University, Hangzhou 310027, China}
\affiliation{International Joint Innovation Center, Zhejiang University, Haining 314400, China}

\begin{abstract}
\textbf{Complete quantum control of a stationary quantum bit embedded in a quantum emitter is crucial for photonic quantum information technologies. Recently, the orbital degree of freedom in optically active quantum dots has emerged as a promising candidate. However, the essential ability to perform arbitrary rotations on orbital qubits remains elusive. Here, we demonstrate arbitrary rotation of a hole orbital qubit with direct phase control using picosecond optical pulses. This is achieved by successfully inducing stimulated Raman transitions within $\Lambda$ systems coupled via radiative Auger processes. The new capability enables direct control of polar and azimuth angles of the Bloch vector without requiring timed precession. Our results establish orbital states in solid-state quantum emitters as a viable resource for applications in high-speed quantum information processing.} 
\end{abstract}

\maketitle

\begin{figure*}
\refstepcounter{fig}
	\includegraphics[width=0.95\linewidth]{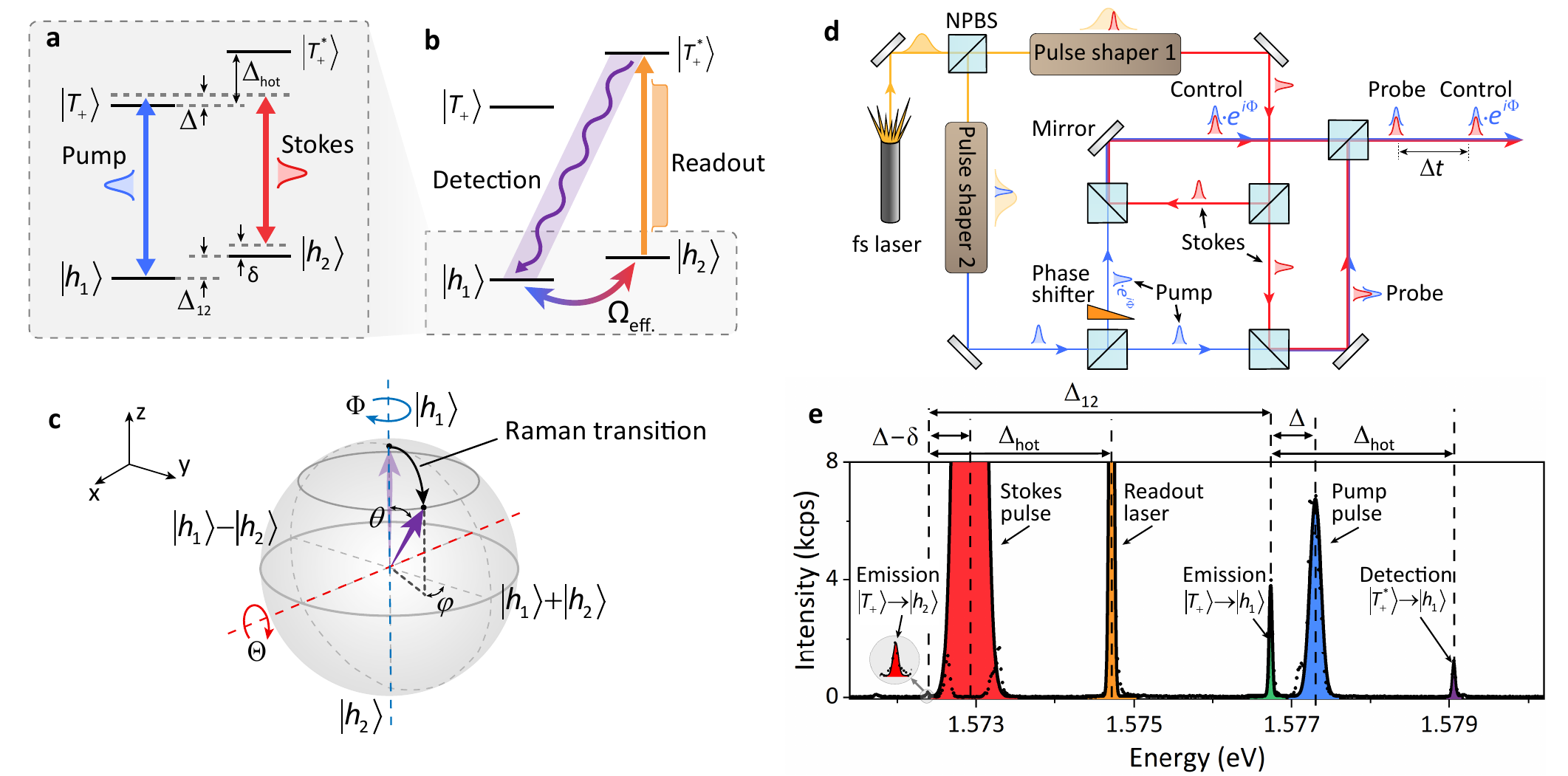}
	\caption{\textbf{Schematic of the phase-controlled stimulated Raman transition.} \textbf{a}, Energy level diagram for Raman transition. Temporally overlapping pump and Stokes pulses result in an effective coupling $\Omega_{\rm{eff.}}$ between hole orbital states. \textbf{b}, Readout scheme for $\left|h_2\right>$. A CW laser (orange) resonant with $\left|h_2\right>\leftrightarrow\left|T_+^*\right>$ transition populates $\left|T_+^*\right>$, leading to fluorescence signal (purple) proportional to the $\left|h_2\right>$ population. \textbf{c}, Bloch sphere representation of optically induced rotation of a orbital qubit. The qubit can be driven to an arbitrary point on the Bloch sphere by the Raman pulse. The polar angle, $\theta$, can be controlled by the pulse area $\Theta$. The azimuth angle, $\varphi$, can be set by the initial phase difference $\Phi$ between pump and Stokes pulses. \textbf{d}, Experimental setup for generating a pair of phase-controlled Raman pulses (control and probe). fs laser: femtosecond laser. NPBS: non-polarizing beam splitter. \textbf{e}, Measured fluorescence and laser spectrum. Black lines: Gaussian fits. A dip in the center of the Stokes pulse spectrum arises from a notch filter. The energy difference between the fundamental transition ($\left|T_+\right>\leftrightarrow\left|h_1\right>$) and radiative Auger transition ($\left|T_+\right>\leftrightarrow\left|h_2\right>$) is $\Delta_{12}= 4.31~\rm{meV}$ (1.04~THz). The pump and Stokes pulses are detuned from the fundamental and radiative Auger transitions by $\Delta=0.57~\rm{meV}$ and $\Delta-\delta=0.52~\rm{meV}$, respectively. kcps: kilo counts per second.}
	\label{Fig1}
\end{figure*}

A stationary qubit interfacing with a flying qubit plays an essential role in quantum information technologies, such as quantum networks~\cite{Ramakrishnan2023,Lu2021b} and quantum computing with photonic cluster states~\cite{Walther2005,Lindner2009,Economou2010a}. This system can be realized by embedding a stationary qubit in an optically active quantum system, e.g., trapped ions~\cite{Blatt2008}, cold atoms~\cite{Thomas2022,Yang2022} and color centers~\cite{Bernien2013}. Among these candidates, solid-state epitaxially grown semiconductor quantum dots (QDs) attract much attention due to their high optical quality~\cite{Zhai2022,Ding2016d,Huber2017a} and compatibility with nanophotonic structures~\cite{Somaschi2016,Liu2018f,Liu2019e,Tiranov2023}. In recent decades, advancements in stationary qubits within optically active QDs have primarily focused on the spin degree of freedom (DoF), including the generation of spin–photon entanglement~\cite{DeGreve2012a,Gao2012}, multiphoton cluster state~\cite{Coste2023,Cogan2023,Appel2022} and spin-spin entanglement~\cite{Delteil2016a}.

Despite great progress with spin qubits, another important DoF of confined carriers inside QDs, namely orbital states, has long been neglected. Orbital states offer great potential for realizing solid-state qubits with high fidelity and nearly lifetime-limited coherence~\cite{Yan2023}. Additionally, manipulating an orbital qubit does not require an external magnetic field, thereby reducing the complexity of the experimental setup. Furthermore, combining orbital and spin degrees of freedom, in theory, allows the realization of a CNOT gate with a single charge carrier~\cite{Monroe1995}. However, unlike well-established optical manipulation techniques for spin qubits~\cite{Press2008,Press2010a,Gao2015}, research on the coherent control of orbital states in optically active QDs remains limited due to the lack of suitable optical methods for driving orbital transitions which typically occur in the terahertz regime. 

Recently, the radiative Auger process observed in single epitaxial QDs opens a new avenue for coherently manipulating orbital states~\cite{Spinnler2021,Lobl2020,Yan2023,Gawarecki2023}. This process has been employed to create a superposition of hole orbital states using a two-step coherent population transfer scheme~\cite{Yan2023}. This scheme, however, lacks direct phase control and requires a tailored optical pulse sequence for each initial orbital state, hindering its application as a universal quantum gate. Moreover, the involvement of the intermediate trion state introduces additional decoherence and leakage of quantum information. Therefore, to realize universal single-orbital-qubit gates, a protocol allowing phase-controlled arbitrary unitary rotations is mandatory but remains unachieved.

In this letter, we demonstrate arbitrary rotation of hole orbital states with direct phase control using a pair of two-color picosecond pulses. This protocol is enabled by successfully inducing stimulated Raman transition (SRT)~\cite{Press2010a,Bodey2019} within $\Lambda$ systems connected via radiative Auger processes. The control of Bloch vector's polar ($\theta$) and azimuth ($\varphi$) angles is verified by Rabi oscillations and Ramsey interference patterns, respectively. Finally, arbitrary rotation of hole orbital states is demonstrated by simultaneously varying $\theta$ and $\varphi$ via scanning the area $\Theta$ and phase $\Phi$ of Raman pulses. In contrast to SRT schemes based on a single broad pulse~\cite{Press2008,Berezovsky2008,Buckley2010}, which require a timed precession to generate arbitrary states, the double-pulse scheme~\cite{Bodey2019, Zhou2017, Sweeney2011a} employed here allows direct control of orbital qubit's phase (the azimuth angle $\varphi$), which is crucial for scalable operations on multiple qubits. Our demonstration lays the foundation for the development of orbital qubits in solid-state quantum emitters and could potentially be extended to gate-defined QDs, enabling ultrafast quantum gates for charge qubits~\cite{Petta2004,Cao2013,Descamps2023,Fujita2019}.

The experiments are performed on a single GaAs/AlGaAs QD grown using local droplet etching technique \cite{Babin2022,Stemmann2008a} and embedded in an \textit{n-i-p} diode device~\cite{Zhai2020,Schimpf2021,Warburton2000}. The QD sample is held at 3.6~K and the fluorescence is collected using a confocal microscope. The energy level diagrams of the QD, depicted in Figs.~\ref{Fig1}a and b, show the ground and excited states of a trapped hole ($\left|h_1\right>$ and $\left|h_2\right>$ split by $\Delta_{\rm{12}}$) and a positive trion ($\left|T_+\right>$ and $\left|T_+^*\right>$ split by $\Delta_{\rm{hot}}$). $\left|T_+\right>$ consists of two holes in the lowest orbital $h_1$ and one electron, whereas $\left|T_+^*\right>$ includes a hole in each of $h_1$ and $h_2$ orbitals plus a ground-state electron (see Supplementary Figs.~1~\cite{SI}). Both trion states are optically connected to $\left|h_1\right>$ and $\left|h_2\right>$ via fundamental transition and radiative Auger transition, forming two independent $\Lambda$ systems. 

\begin{figure*}
\refstepcounter{fig}
	\includegraphics[width=1\linewidth]{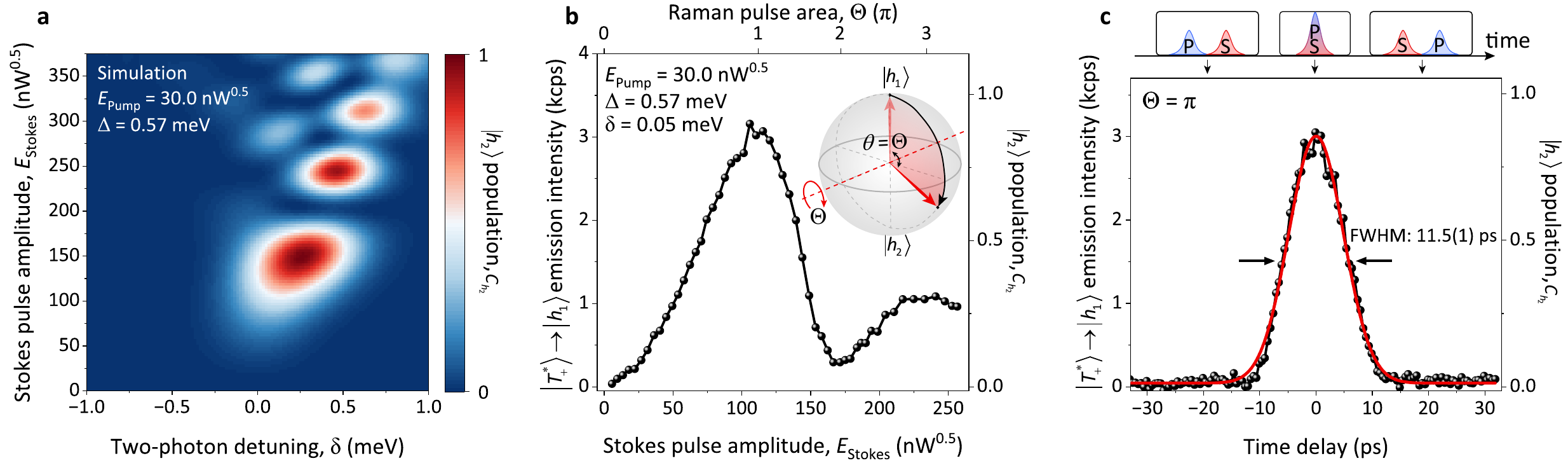}
	\caption{\textbf{Rabi oscillation and control of polar angle $\theta$.} \textbf{a}, Simulated $\left|h_2\right>$ population as a function of Stokes pulse amplitude $E_{\rm{Stokes}}$ and two-photon detuning $\delta$, where $\Delta$, $E_{\rm{Pump}}$ are fixed at 0.57~meV and 30.0~${\rm{nW}}^{0.5}$. \textbf{b}, Rabi oscillation. $\left|T_+^*\right>\rightarrow\left|h_1\right>$ emission intensity, proportional to the $\left|h_2\right>$ population, as a function of the Stokes pulse amplitude. Inset: Example of a trajectory on the Bloch sphere. \textbf{c}, $\left|T_+^*\right>\rightarrow\left|h_1\right>$ emission intensity as a function of time delay between pump and Stokes pulses. Red: a Gaussian fit. P: pump pulse. S: Stokes pulse.}
	\label{Fig2-Rabi}
\end{figure*}
\begin{figure*}
\refstepcounter{fig}
	\includegraphics[width=1\linewidth]{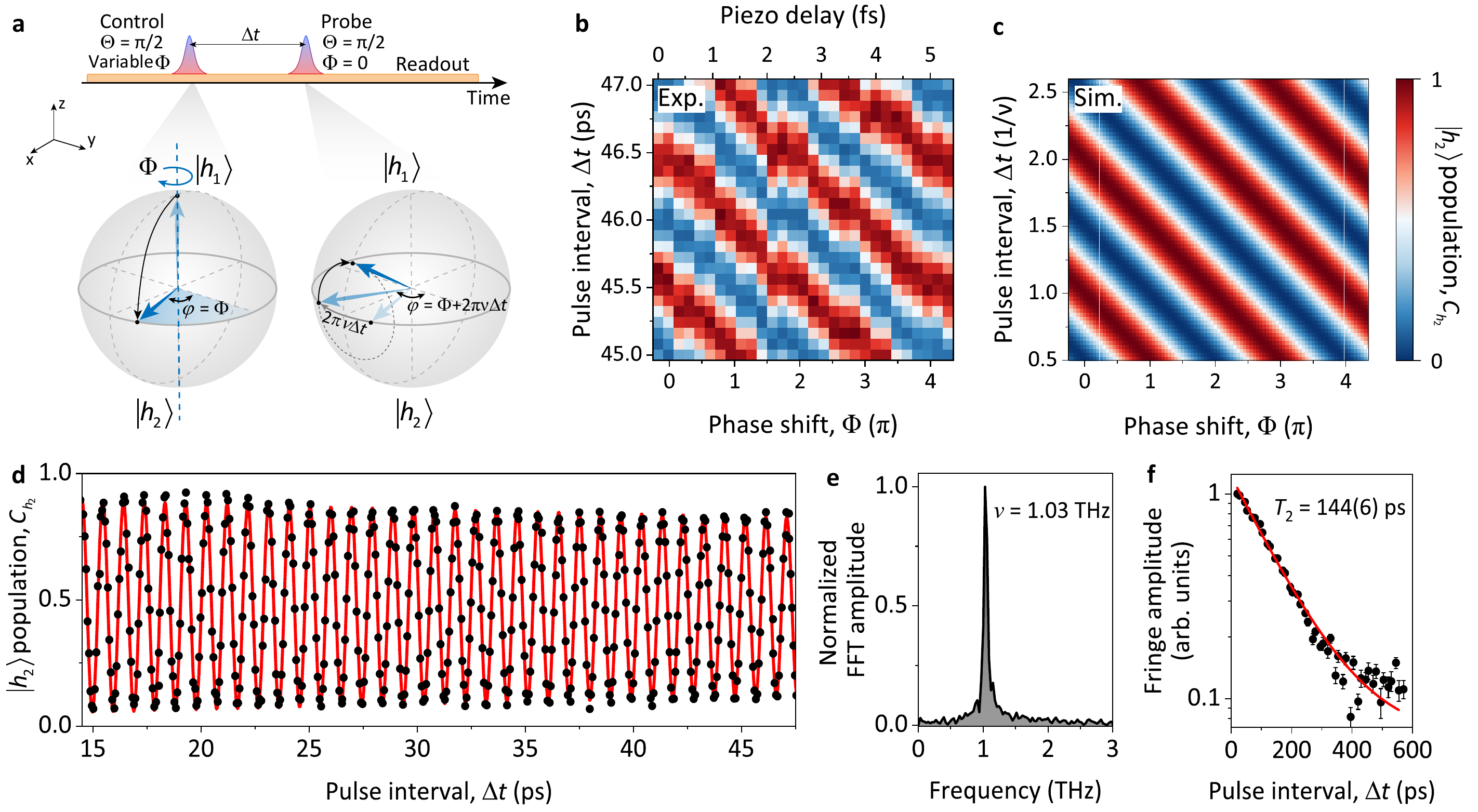}
	\caption{\textbf{Phase-controlled Ramsey interference and manipulation of azimuth angle $\varphi$.} \textbf{a}, Pulse sequence and corresponding vector trajectory. Left: the Bloch vector is driven to the equator by the first $\pi/2$ Raman pulse (control) with the phase $\Phi$. Right: the resulting state is probed by the second $\pi/2$ Raman pulse (probe) after a variable pulse interval $\Delta t$. \textbf{b},~Experimentally recorded $\left|h_2\right>$ population as functions of the relative phase of the control pulse and the pulse interval between the control and probe pulses. Continuously shifting interference fringes demonstrate the control of the Bloch vector's azimuth angle. \textbf{c},~The simulation obtained from a simplified two-level model. \textbf{d}, A Ramsey fringe in a larger time interval span. Red: a sinusoidal fit. \textbf{e}, Fast Fourier transform (FFT) of Ramsey fringe. \textbf{f}, The fringes amplitude as a function of pulse interval. Red: a single-exponential fit.}
	\label{Fig3}
\end{figure*}

To achieve arbitrary rotation of the orbital qubit consisting of $\left|h_1\right>$ and $\left|h_2\right>$ (Fig.~\ref{Fig1}c), we induce SRT within the $\Lambda$ system linked by $\left|T_+\right>$ (Figs.~\ref{Fig1}a). This is accomplished using a pair of temporally overlapping pump and Stokes pulses generated from the setup shown in Fig.~\ref{Fig1}d. The phase-locked pump and Stokes pulses are combined, and the resulting pulse is referred to as a Raman pulse in the rest of the paper. The population of $\left|h_2\right>$ generated by the Raman process (denoted as $C_{h_2}$) can be read out by resonantly driving the $\left|h_2\right> \rightarrow \left|T_+^*\right>$ transition using a weak continuous-wave (CW) laser (Fig.~\ref{Fig1}b, orange) and comparing the $\left|T_+^*\right>\rightarrow\left|h_1\right>$ emission intensity (purple) detected with and without the CW laser. The measured spectrum of optical transitions and laser pulses involved in the SRT is shown in Fig.~\ref{Fig1}e.

\begin{figure*}
\refstepcounter{fig}
	\includegraphics[width=1\linewidth]{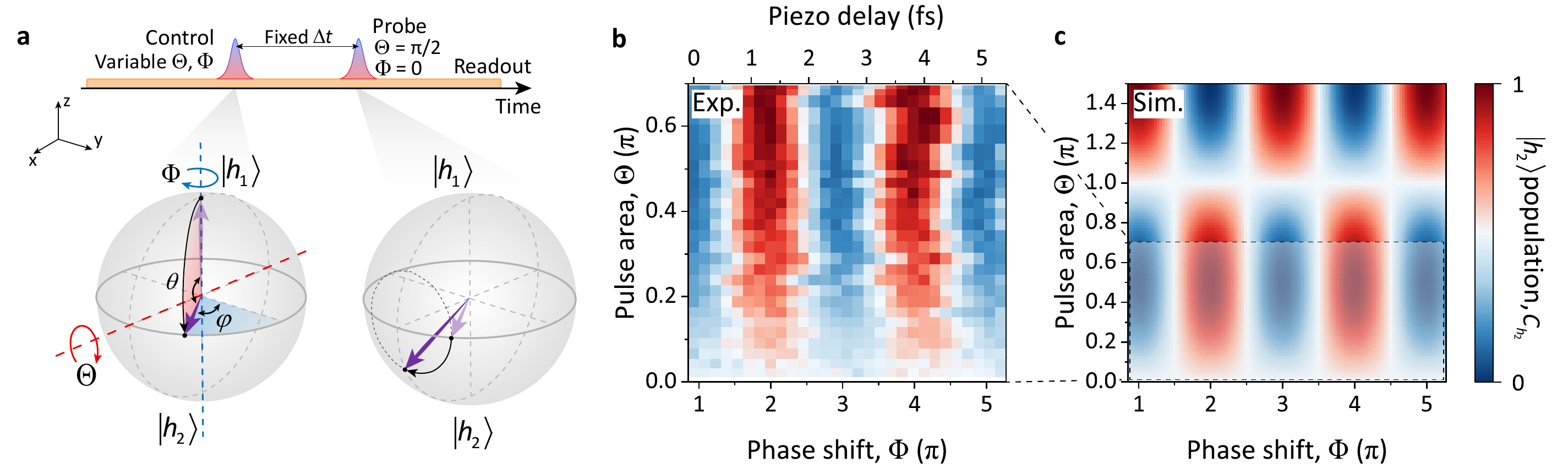}
    \caption{\textbf{Simultaneous control of polar and azimuth angles.} \textbf{a}, Pulse sequence and corresponding vector trajectory. Left: the Bloch vector is driven to an arbitrary point on the Bloch sphere by the control pulse with variable pulse area and phase. Right: the resulting state is probed by a subsequent $\pi/2$ Raman pulse (probe) with a fixed pulse interval $\Delta t=n/\nu$, $n$ is an integer number. \textbf{b},~Experimentally recorded $\left|h_2\right>$ population as functions of $\Phi$ and $\Theta$ of the control pulse for a fixed pulse interval. \textbf{c},~The simulation covering a larger $\Theta$ range.
    }
	\label{Fig4}
\end{figure*}

For an ideal stimulated Raman process~\cite{Bodey2019,Zhou2017,Sweeney2011a}, pump and Stokes pulses separately drive the fundamental and radiative Auger transitions with Rabi frequencies of $\Omega_{\rm{pump}}$, $\Omega_{\rm{Stokes}}$ and a single-photon detuning of $\Delta$ (Fig.~\ref{Fig1}a). In the case of $\Delta\gg\Omega_{\rm{pump}}$ and $\Delta\gg\Omega_{\rm{Stokes}}$, the system can be simplified as an effective two-level system by adiabatically eliminating $\left|T_+\right>$. The Bloch vector, determined by its polar~($\theta$) and azimuth~($\varphi$) angles, can then be independently rotated around the x-axis by the effective pulse area $\Theta=\int_{-\infty}^{\infty}\Omega(t) dt$ and around the z-axis by the initial phase difference $\Phi$ between pump and Stokes pulses~\cite{Sweeney2011a,Bodey2019} (Figs.~\ref{Fig1}b~and~c). $\Omega(t)\propto\Omega_{\rm{pump}}\Omega_{\rm{Stokes}}/{\Delta}$ is the effective Rabi frequency~\cite{Press2008}. 

However, in our system, the situation becomes more complicated because each pulse simultaneously interacts with both transitions, and there is a significant difference in dipole moments between these two types of transitions~\cite{SI}. These combined factors result in an unbalanced optical Stark shift, hindering the Raman process. To address this problem, we introduce a static two-photon detuning $\delta$ which essentially compensates for the unbalanced optical Stark shift~\cite{Tinkey2022}. The master equation simulation (see Supplementary~Material~Section~III~\cite{SI}) reveals that a near-unity (0.985) population transfer can be achieved~(see Fig.~\ref{Fig2-Rabi}a). In addition, this issue may also be resolved with a recently proposed protocol utilizing chirped pulses~\cite{Chathanathil2023}.

Following the optimization of the coherent control protocol, we first present experimental evidence of control over the Bloch vector's polar angle $\theta$. The qubit is initially at $\left|h_1\right>$. The Raman pulse with a pulse area of $\Theta$ rotates the Bloch vector, resulting an $h_2$ population of $C_{h_2}$=sin$^2$$(\theta/2)$ and $\theta=\Theta$. Figure~\ref{Fig2-Rabi}b shows the measured $C_{h_2}$ as a function of the $E_{\rm{Stokes}}$ with $E_{\rm{pump}}$ fixed at 30.0~$\rm{nW^{0.5}}$ (effectively, this corresponds to a pulse area of 1.94$\pi$ for the fundamental transition). The Rabi oscillation of $C_{h_2}$ demonstrates changes in $\theta$ up to $3\pi$. The fidelity of the $\pi$-rotation is estimated to be $87.3\%$ (Supplementary Section~IV~\cite{SI}).

To verify that the population transfer to $\left|h_2\right>$ observed above indeed occurs via the SRT, we investigate the dependence of $C_{h_2}$ on the time delay between pump and Stokes pulses with a fixed field amplitude corresponding to $\Theta=\pi$ (Fig.~\ref{Fig2-Rabi}c). The temporal profile of $C_{h_2}$ displays a width of 11.5(1)~ps, matching the width of the two-pulse convolution derived from the single-pulse duration ($\sim$8.49~ps). This result confirms that $\left|h_2\right>$ can be efficiently populated solely when two pulses overlap temporarily, a clear signature of stimulated Raman processes.

Next, we move to the demonstration of control over the azimuth angles $\varphi$ of the Bloch vector via the phase-controlled Ramsey interference. The measurement procedure is illustrated in Fig.~\ref{Fig3}a where the QD is driven by a pair of $\pi/2$ Raman pulses, referred to as control and probe, with a variable pulse interval $\Delta t$. Unlike standard Ramsey interference~\cite{Greilich2011,Godden2012}, here, the optical phase of the control pulse can be adjusted by a phase shifter (Fig.~\ref{Fig1}d). This extra flexibility enables direct manipulation over the initial azimuth angle. The $\pi/2$ control pulse rotates the Bloch vector to the equator with the azimuth angle $\varphi=\Phi$, creating a superposition state $\left|\Psi\right>=\frac{1}{\sqrt{2}}(\left|h_1\right>+e^{i\Phi}\left|h_2\right>)$. Then the qubit undergoes a free precession with the frequency $\nu=\Delta_{12}/h$, leading to $\varphi=\Phi+2\pi\nu\Delta t$. Here, pure dephasing is neglected. A detailed simulation that includes experimental noise and system dephasing can be found in Supplementary Section~III~\cite{SI}. The subsequent $\pi/2$ probe pulse drives the Bloch vector towards $\left|h_2\right>$ or $\left|h_1\right>$ depending on both the initial phase $\Phi$ and phase accumulation $2\pi\nu\Delta t$ within the pulse interval. We measure the population of $\left|h_2\right>$ as a function of $\Delta t$ at different $\Phi$ of the control pulse (Fig.~\ref{Fig3}b). As we sweep $\Phi$, the phase of the Ramsey fringe shifts, while the fringe amplitude remains constant. For comparison, we show the master-equation simulation for a simplified two-level model (Fig.~\ref{Fig3}c). The good agreement confirms the successful mapping of the control pulse's $\Phi$ to the azimuth angle $\varphi$. 

Figure~\ref{Fig3}d presents a representative fringe in a larger time interval span at a constant $\Phi$. The Fourier transform reveals a peak oscillation frequency of 1.03~THz (Fig.~\ref{Fig3}e), aligning well with $\Delta_{12}$ observed in the PL spectrum (Fig.~\ref{Fig1}e). To evaluate the coherence time, $T_2$, we plot the fringe amplitudes for a series of coarse delays (Fig.~\ref{Fig3}f). The fringe amplitude decreases as $\Delta t$ increases. The decay is best fitted to an exponential function (red), from which $T_2=144(6)$~ps is obtained. This allows us performing dozens of single-qubit operations within the coherence time. We note that the relatively short $T_2$ compared with $2T_1$ (318(4)~ps, Supplementary Fig.~9~\cite{SI}) is not limited by the intrinsic coherence property of orbital qubit, but mainly due to the presence of the readout CW laser which acts as an additional dephasing channel (see Supplementary Fig.~2~\cite{SI,Press2008}). 

After separately verifying the ability to control the polar and azimuth angles of the qubit, we now proceed to demonstrate arbitrary rotation by simultaneously adjusting both angles. In Fig.~\ref{Fig4}a, we fix $\Delta t$ and sweep up the $\Theta$ of the control pulse across various $\Phi$. The resulting state is then rotated by the subsequent $\pi/2$ probe pulse. The measured population of $\left|h_2\right>$ exhibits clear interference fringes as shown in Fig.~\ref{Fig4}b. This pattern is well reproduced by our master-equation simulation describing a Bloch vector with continuously varying $\Theta$ and $\Phi$ interfered with a $\pi/2$ pulse (Fig.~\ref{Fig4}c). This good agreement confirms our ability to simultaneously manipulate the Bloch vector's polar and azimuth angles, demonstrating phase-controlled arbitrary rotation of the hole orbital qubit. This SRT protocol can also be extended to manipulate other higher orbital states (see Supplementary Section~III~\cite{SI}), suggesting the potential for high-dimensional quantum information processing~\cite{Ringbauer2021, Chi2022}.

In conclusion, we successfully induce a stimulated Raman transition in a $\Lambda$ system linked via radiative Auger processes, which fundamentally differs from conventional dipole-allowed optical transitions~\cite{Press2010a,Bodey2019}. This enables ultrafast arbitrary rotation of a hole orbital qubit with direct phase control in an optically active QD. Additionally, since intermediate states are nearly adiabatically eliminated in SRT, our approach avoids additional dephasing and allows unitary operations for the two-level system composed of two hole-orbital states, which is essential for implementing universal quantum gates. Our work advances orbital-based quantum photonic devices, including the generation of time-bin multiphoton graph state~\cite{Tiurev2021,Appel2022}, orbital-frequency entanglement and exploration of non-Hermitian physics~\cite{Wu2019}. Moreover, Our approach holds significant potential for applications across colloidal nanostructures~\cite{Antolinez2019,Llusar2020}, donor/acceptor-bound excitons~\cite{Dean1967,Bryja2016} and quantum emitters in two-dimensional materials~\cite{Seyler2019,Alexeev2019,Baek2020,Binder2019}. 

\newpage
We thank Dr. Xin Zhang for fruitful discussions. We also thank Nadine Viteritti for the electrical contact preparation of the sample. F.L. acknowledge support from the National Key Research and Development Program of China (2023YFB2806000, 2022YFA1204700), National Natural Science Foundation of China (U21A6006, 62075194). L.Z. acknowledge support from SNF Project 200020\_204069. H.-G.B., A.D.W., and A.L. acknowledge support by the BMBF-QR.X Project 16KISQ009 and the DFH/UFA, Project CDFA-05-06.

\addtocontents{toc}{\SkipTocEntry}
\putbib 
\end{bibunit}

\begin{bibunit}[naturemag]

\setcounter{equation}{0}
\setcounter{figure}{0}
\setcounter{table}{0}

\renewcommand\thesection{\Roman{section}}
\renewcommand{\figurename}{Supplementary Fig.}
\renewcommand{\tablename}{Supplementary Table}
\renewcommand{\thetable}{\arabic{table}}

\renewcommand{\theequation}{S\arabic{equation}}
\renewcommand{\bibnumfmt}[1]{[S#1]} 
\renewcommand{\citenumfont}[1]{S#1} 

\onecolumngrid
\newpage

\begin{Large}
\begin{center}
\textbf{Supplementary Material: All-optical ultrafast arbitrary rotation of hole orbital qubits with direct phase control}
\end{center}
\end{Large}

\setcounter{page}{1}

\begin{center}
Jun-Yong Yan, Liang Zhai, Hans-Georg Babin, Yuanzhen Li, Si-Hui Pei, Moritz Cygorek, Wei Fang, Fei Gao,\\ Andreas D. Wieck, Arne Ludwig, Chao-Yuan Jin, Da-Wei Wang, and Feng Liu$^*$
\end{center}

\tableofcontents

\newpage
\section{Identification of the hot trion state and evaluation of the readout probability}
\label{Section-indentification}

To identify the hot trion state, we measure the fluorescence spectra under resonant excitation of $\left|h_1\right>\rightarrow\left|T_+\right>$ and $\left|h_1\right>\rightarrow\left|T_+^*\right>$ transitions by a narrow-band CW laser. As shown in Supplementary Fig.~\ref{Sfig:identify}, Auger peaks originating from both states $\left|T_+\right>$ and $\left|T_+^*\right>$ exhibit the same energy difference (indicated by the gray dashed lines). This observation suggests that the final states involved in the Auger emission process are identical for both states. Furthermore, when resonantly exciting $\left|T_+^*\right>$, we also observed $\left|T_+\right>\rightarrow\left|h_1\right>$ emission. This observation suggests that $\left|T_+^*\right>$ can relax to $\left|T_+\right>$. Based on these results, we conclude that $\left|T_+^*\right>$ is a hot trion state that consists of two holes (one in the ground state and another in the first excited state) and one electron (in the ground state) as shown in the right inset of Supplementary Fig.~\ref{Sfig:identify}d.

\begin{figure}[h]
\includegraphics[width=1\textwidth]{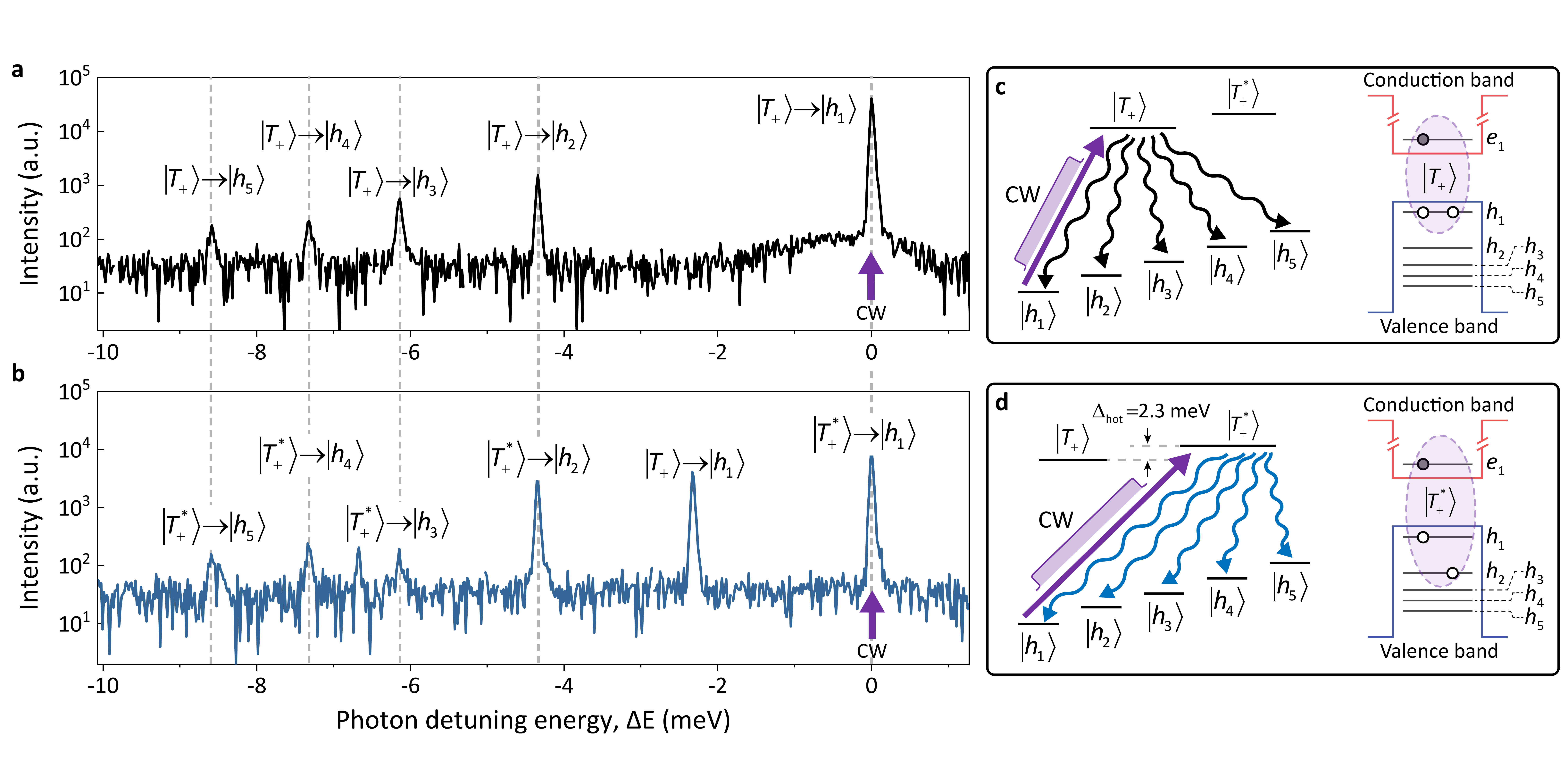}
\caption{\textbf{a}, Fluorescence spectrum under resonant excitation of $\left|h_1\right>\rightarrow\left|T_+\right>$ transition (purple arrow). The highest peak originates from the fundamental transition ($\left|T_+\right>\rightarrow\left|h_1\right>$) and a series of peaks at the lower energy side correspond to radiative Auger transitions from $\left|T_+\right>$ to high-orbital hole states. \textbf{b}, Fluorescence spectrum under resonant excitation of  $\left|h_1\right>\rightarrow\left|T_+^*\right>$ transition (purple arrow). The highest energy peak comes from the fundamental transition ($\left|T_+^*\right>\rightarrow\left|h_1\right>$). A series of peaks at the lower energy side correspond to radiative Auger transitions from $\left|T_+^*\right>$ to high-orbital hole states. The gray dahsed lines are guides to the eye, indicating that the energy difference between Auger peaks and fundamental peak is the same in \textbf{a} and \textbf{b}. \textbf{c} (\textbf{d}), Schematic description of resonant excitation, fundamental emission and Auger emission processes of $\left|T_+\right>$ ($\left|T_+^*\right>$). Right: Energy-level diagram of $\left|T_+\right>$ ($\left|T_+^*\right>$).}
\label{Sfig:identify}
\end{figure}

\begin{figure}[h]
\includegraphics[width=0.95\textwidth]{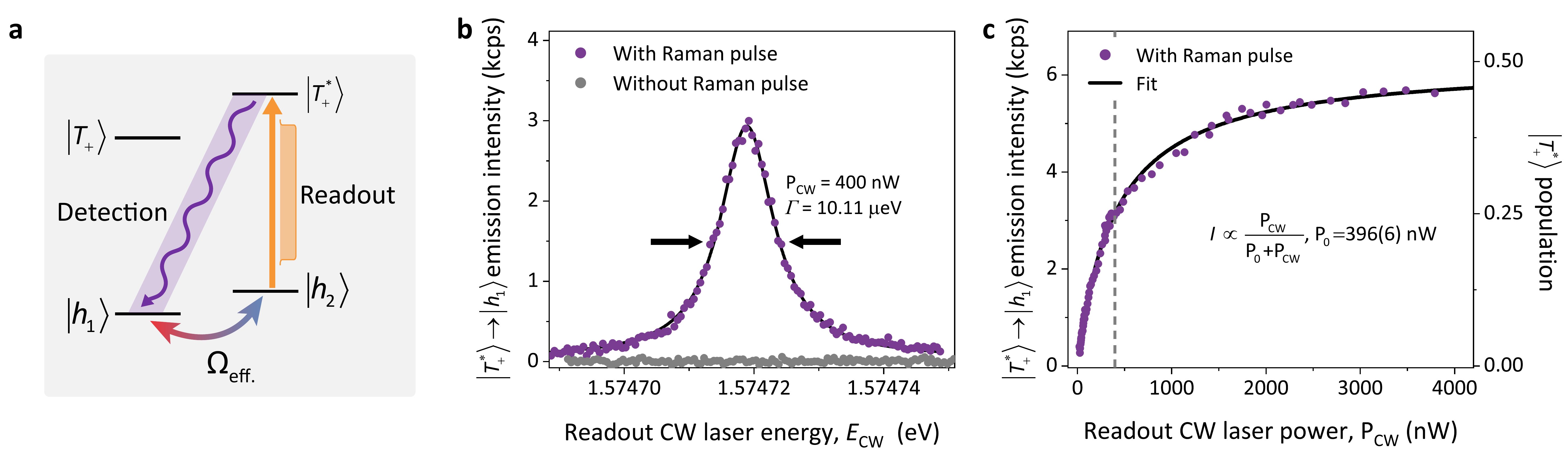}
\caption{\textbf{a}, Energy level scheme of the double $\Lambda$ system. \textbf{b}, Detected intensity as a function of the energy of readout CW laser with (purple) and without (gray) the Raman pulse. The power of the CW laser is 400~nW. The readout CW laser will not cause any fluorescence in the absence of Raman pulse, because there is no population in the $h_2$ orbital. Black curve: a Lorentzian fit to the data with a linewidth $\it{\Gamma}$ $=10.11~\mu\rm{eV}$. \textbf{c}, Detected intensity as a function of readout CW laser power $\rm{P}_{CW}$ with Raman pulse. Black curve: a fitting using the function $I\propto\rm{P}_{CW}/(\rm{P}_{0}+\rm{P}_{CW})$~\cite{Nguyen2012a} with the saturation power $\rm{P}_{0}=396(6)~nW$ (dashed line).}
\label{Sfig:cw scan linewidth}
\end{figure}

The population of the orbital state is read out by tuning a CW laser to resonance with the $\left|h_2\right>\leftrightarrow\left|T_+^*\right>$ transition (orange arrow in Fig.~\ref{Sfig:cw scan linewidth}a). To maximize the readout probability, we scan the CW laser energy ($E_{\rm{CW}}$) while monitoring the $\left|T_+^*\right>\rightarrow\left|h_1\right>$ emission. The results are shown in Supplementary Fig.~\ref{Sfig:cw scan linewidth}b, where the emission intensity is plotted as a function of $E_{\rm{CW}}$. From this result, we determine the resonance energy to be 1.57472~eV and the linewidth to be $10.11~\mu\rm{eV}$.

Next, we investigate the dependence of the emission intensity on the CW laser power ($\rm{P}_{\rm{CW}}$), as shown in Supplementary Fig.~\ref{Sfig:cw scan linewidth}c. The observed power dependence aligns well with the theoretical power saturation curve for a two-level system ($I\propto\rm{P}_{CW}/(\rm{P}_{0}+\rm{P}_{CW})$~\cite{Nguyen2012a} with saturation power $\rm{P}_{0}=396(6)~nW$ (dashed line)). We choose a readout CW laser of 400~nW, corresponding a readout probability of $\sim25\%$. It is important to note that the coupling between $\left|h_2\right>$ and $\left|T_+^*\right>$ induced by the readout CW laser can potentially deteriorate the measured coherence time $T_2$ from Ramsey interference (Fig.~3f). This limitation could be addressed in future experiments by switching the CW laser off between two Raman pulses using a fast electro-optic modulator or employing a picosecond readout pulse. 

\newpage

\section{Dipole moments of single-photon transitions}
\label{Section-dipole}

\begin{figure}[h]
\includegraphics[width=1\textwidth]{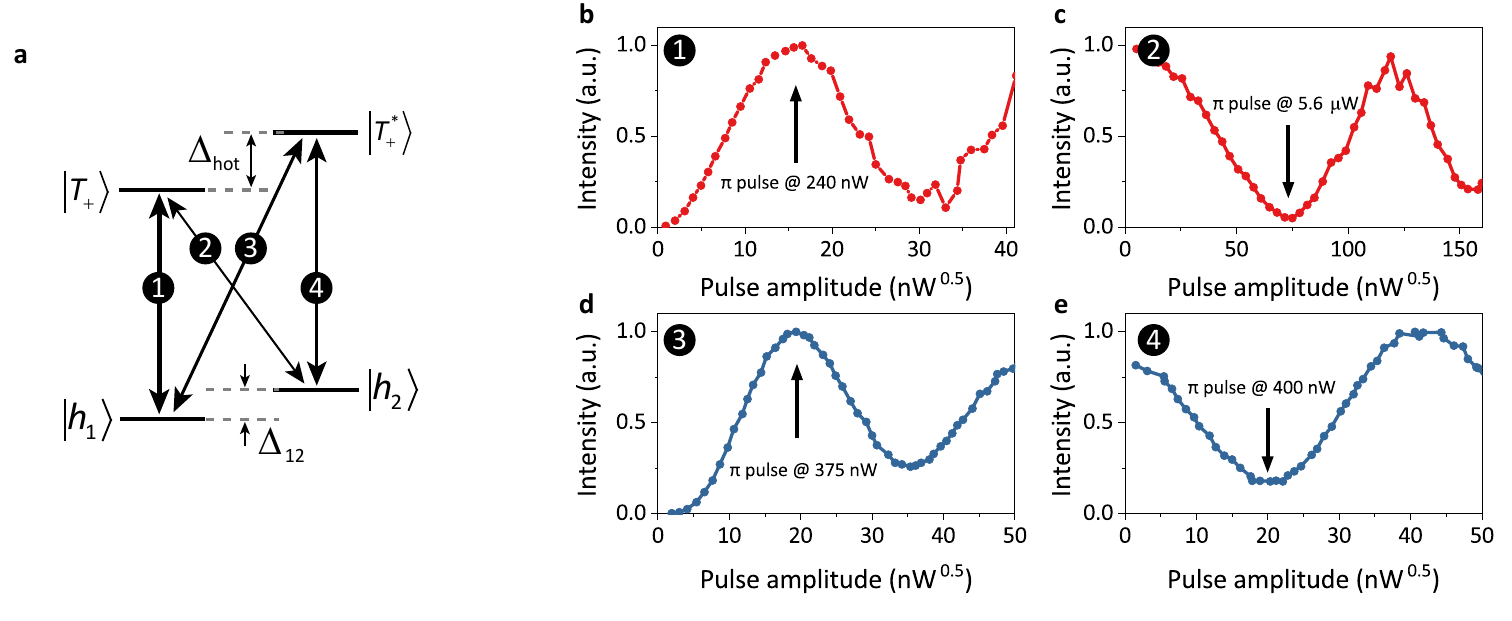}
\caption{\textbf{a}, Level scheme of transitions involved in double $\Lambda$ system. \textbf{b-e}, Rabi oscillations. By resonantly driving Rabi oscillations of each single-photon transition, we extracted the dipole moment ratio of four transitions: $\mu_1:\mu_2:\mu_3:\mu_4=1:\frac{1}{4.8}:\frac{1}{1.25}:\frac{1}{1.29}$.}
\label{Sfig:dipole moment}
\end{figure}

To obtain the dipole moments involved in our experiment (Supplementary Fig.~\ref{Sfig:dipole moment}a), we measure the Rabi oscillations of each single-photon transition. 
For transitions 1 and 3, we directly extract the $\pi$ pulse power where the emission intensity achieves the first maximum. For transition 2 (4), we first pump the population to $\left|T_+\right>$ ($\left|T_+^*\right>$), then introduce another pulse to depopulate the population to $\left|h_2\right>$, and extract the $\pi$ pulse power from the first dip of the oscillation. From the ratio of $\pi$ pulse power, we determine the dipole moment ratio of four transitions
\begin{equation}
\mu_1:\mu_2:\mu_3:\mu_4=1:\frac{1}{4.8}:\frac{1}{1.25}:\frac{1}{1.29}.
\end{equation}

\newpage
\section{Modeling}
\label{section: Theoretical modeling}
The four-level system is described by the following Hamiltonian with basis \{$\left|h_1\right>$, $\left|T_+\right>$, $\left|T_+^*\right>$, $\left|h_2\right>$\} under rotating wave approximation:

\begin{equation}
H_{\text {Lab.}} / \hbar=\left(\begin{array}{cccc}
0 & \frac{1}{2} \Omega_{P 1} e^{i \omega_P t}+\frac{1}{2} \Omega_{S 1} e^{i \omega_S t} & \frac{1}{2} \Omega_{P 3} e^{i \omega_P t}+\frac{1}{2} \Omega_{S 3} e^{i \omega_S t} & 0 \\
\frac{1}{2} \Omega_{P 1} e^{-i \omega_P t}+\frac{1}{2} \Omega_{S 1} e^{-i \omega_S t} & \omega_t & 0 & \frac{1}{2} \Omega_{P 2} e^{-i \omega_P t}+\frac{1}{2} \Omega_{S 2} e^{-i \omega_S t} \\
\frac{1}{2} \Omega_{P 3} e^{-i \omega_P t}+\frac{1}{2} \Omega_{S 3} e^{-i \omega_S t} & 0 & \omega_t+\Delta_{\text{hot}} & \frac{1}{2} \Omega_{P 4} e^{-i \omega_P t}+\frac{1}{2} \Omega_{S 4} e^{-i \omega_S t} \\
0 & \frac{1}{2} \Omega_{P 2} e^{i \omega_P t}+\frac{1}{2} \Omega_{S 2} e^{i \omega_S t} & \frac{1}{2} \Omega_{P 4} e^{i \omega_P t}+\frac{1}{2} \Omega_{S 4} e^{i \omega_S t} & \Delta_{12}
\end{array}\right).
\end{equation}

$\Omega_{P i}=E_P \mu_i $ ($\Omega_{S i}= E_S\mu_i$) is the Rabi frequency of transition $i$ induced by pump (Stokes) pulse, where $E_P$ ($E_S$) is the time-dependent field amplitude of pump (Stokes) pulse and $\mu_i$ denotes dipole moment of transition $i$ for $i=1,2,3,4$. $\omega_{P}$ ($\omega_{S}$) is the frequency of pump (Stokes) pulse. $\omega_{t}$ is the frequency of $\left|T_+\right>$. Using a rotation operator

\begin{equation}
U_0=\left(\begin{array}{cccc}
e^{-i\left(\omega_t-\omega_P\right) t} & 0 & 0 & 0 \\
0 & e^{-i \omega_t t} & 0 & 0 \\
0 & 0 & e^{-i \omega_t t} & 0 \\
0 & 0 & 0 & e^{-i\left(\omega_t-\omega_S\right) t}
\end{array}\right),
\end{equation}

we obtain the Hamiltonian in rotating frame:

\begin{equation}
H_{\text{Rot.}} / \hbar=U_0^{\dagger} H_{\text{Lab.}} U_0 / \hbar+i \frac{d U_0^{\dagger}}{d t} U_0
\end{equation}

\begin{equation}
H_{\text {Rot.}} / \hbar=\left(\begin{array}{cccc}
\Delta & \frac{1}{2} \Omega_{P 1^{+}}\frac{1}{2} \Omega_{S 1} e^{-i\left(\Delta_{12}+\delta\right) t} & \frac{1}{2} \Omega_{P 3}+\frac{1}{2} \Omega_{S 3} e^{-i\left(\Delta_{12}+\delta\right) t} & 0 \\
\frac{1}{2} \Omega_{P 1^{+}}\frac{1}{2} \Omega_{S 1} e^{i\left(\Delta_{12}+\delta\right) t} & 0 & 0 & \frac{1}{2} \Omega_{P 2} e^{-i\left(\Delta_{12}+\delta\right) t}+\frac{1}{2} \Omega_{S 2} \\
\frac{1}{2} \Omega_{P 3}+\frac{1}{2} \Omega_{S 3} e^{i\left(\Delta_{12}+\delta\right) t} & 0 & \Delta_{\text{hot}} & \frac{1}{2} \Omega_{P 4} e^{-i\left(\Delta_{12}+\delta\right) t}+\frac{1}{2} \Omega_{S 4} \\
0 & \frac{1}{2} \Omega_{P 2} e^{i\left(\Delta_{12}+\delta\right) t}+\frac{1}{2} \Omega_{S 2} & \frac{1}{2} \Omega_{P 4} e^{i\left(\Delta_{12}+\delta\right) t}+\frac{1}{2} \Omega_{S 4} & -\delta+\Delta
\end{array}\right).
\end{equation}

To evaluate the $\left|h_2\right>$ population after the control pulse, that corresponds to the element $\rho_{44}(\textit{t})$ of the $4\times4$ density matrix $\rho(\textit{t})$, we solve the master equation numerically with the help of the Quantum Toolbox in Python (QuTiP)~\cite{Johansson2012a}:
\begin{equation}
i\hbar\frac{d\rho(\textit{t})}{d\textit{t}}=[\textit{H}_{\rm{Rot.}}, \rho(\textit{t})].
\label{Eq: ME1}
\end{equation}

\begin{figure}[h]
\refstepcounter{fig}
	\includegraphics[width=0.9\linewidth]{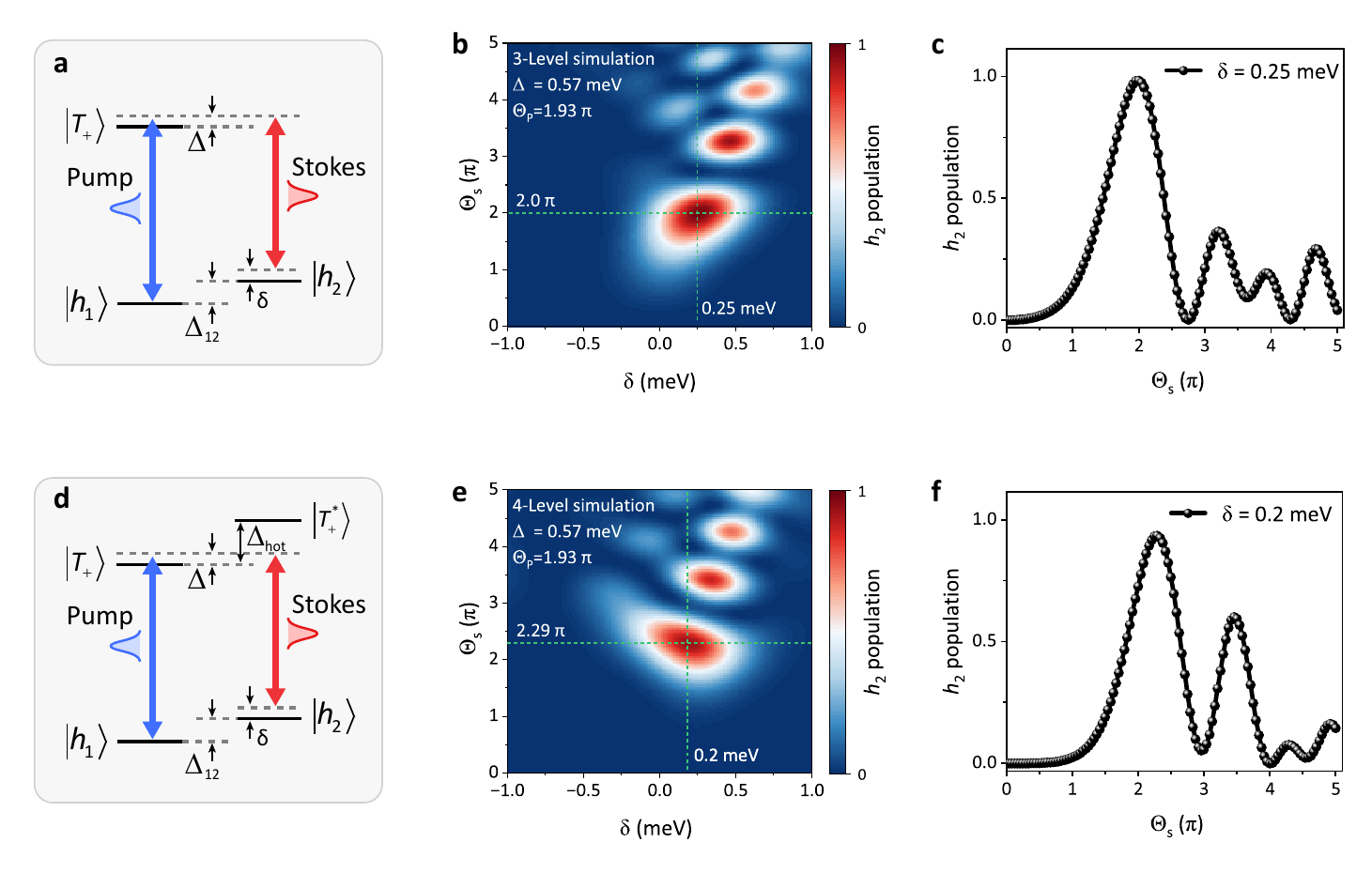}
	\caption{Master-equation simulations without (\textbf{a}, \textbf{b} and \textbf{c}) and with (\textbf{d}, \textbf{e} and \textbf{f}) considering the laser coupling with $\left|T_+^*\right>$.}

\label{Sfig:Master simulation}
\end{figure}

We fix the single-photon detuning $\Delta$ to 0.57 meV and the pump pulse area $\Theta_{\rm{P}}=\int_{-\infty}^{\infty}\Omega_{P1}(t) dt$ to 1.93 $\pi$, as actually used in experiment (Fig.~2). Supplementary Fig.~\ref{Sfig:Master simulation} \textbf{b} (\textbf{e}) shows the calculated final $\left|h_2\right>$ population as a function of the two-photon detuning $\delta$ and Stokes pulse area $\Theta_{\rm{S}}$ without (with) considering the coupling with hot trion. For a 3-level situation, a Raman $\pi$ pulse condition appears at $\delta=0.25$ meV and $\Theta_{\rm{S}}=2.0~\pi$, where $\Theta_{\rm{S}}=\int_{-\infty}^{\infty}\Omega_{S2}(t) dt$. For a 4-level situation, a Raman $\pi$ pulse condition appears at $\delta=0.2$ meV and $\Theta_{\rm{S}}=2.29~\pi$. From the simulation results, we found that the contribution of hot trion on SRT is relatively small.

\newpage
\subsection{Stimulated Raman transition for high-dimensional quantum state manipulation}
\label{section: multidimensional}
The multilevel structure within QD naturally formulates a high-dimensional quantum information encoding resource. To show the capability for extending the SRT approach to other higher states, we theoretically show the coherent manipulation between $h_1$ and $h_3$ states. Additionally, transitions between other orbital states are feasible and can be simulated similarly. In the rotating frame and rotating wave approximation, the four-level system is described by the following Hamiltonian with the basis of \{$\left|h_1\right>$, $\left|T_+\right>$, $\left|h_2\right>$, $\left|h_3\right>$\}:

\begin{equation}
H_{\text {Rot.}} / \hbar=\left(\begin{array}{cccc}
\Delta & \frac{1}{2} \Omega_{P 1^{+}}\frac{1}{2} \Omega_{S 1} e^{-i\left(\Delta_{12}+\delta\right) t} & 0 & 0 \\

\frac{1}{2} \Omega_{P 1^{+}}\frac{1}{2} \Omega_{S 1} e^{i\left(\Delta_{12}+\delta\right) t} & 0 & \frac{1}{2} \Omega_{P 2} e^{-i\left(\Delta_{13}+\delta\right) t}+\frac{1}{2} \Omega_{S 2} & \frac{1}{2} \Omega_{P 5} e^{-i\left(\Delta_{13}+\delta\right) t}+\frac{1}{2} \Omega_{S 5} \\

0 & \frac{1}{2} \Omega_{P 2} e^{i\left(\Delta_{13}+\delta\right) t}+\frac{1}{2} \Omega_{S 2} & -\delta+\Delta-\Delta_{23} & 0 \\

0 & \frac{1}{2} \Omega_{P 5} e^{i\left(\Delta_{13}+\delta\right) t}+\frac{1}{2} \Omega_{S 5} & 0 & -\delta+\Delta
\end{array}\right),
\end{equation}

where $\Delta_{ij}$ ($i,j \in \{1, 2, 3\}$) represents the frequency different between $h_i$ and $h_j$ orbital state. $\Omega_{P i}=E_P \mu_i $ ($\Omega_{S i}= E_S\mu_i$) is the Rabi frequency of transition $i$ induced by pump (Stokes) pulse. The involved transitions and pulses are labeled in Supplementary Figs.~\ref{Sfig:Master simulation for h3} \textbf{a} and \textbf{b}. 

We maintain the single-photon detuning $\Delta$ at 0.57 meV and the pump pulse area $\Theta_{\rm{P}}$ at 1.93 $\pi$. Supplementary Fig.~\ref{Sfig:Master simulation for h3} \textbf{c} presents the calculated final $\left|h_3\right>$ population as a function of the two-photon detuning $\delta$ and Stokes pulse area. The Stokes pulse area $\Theta_{\rm{S}}$ is now defined as $\Theta_{\rm{S}}=\int_{-\infty}^{\infty}\Omega_{S5}(t) dt$. Due to a larger dipole moment difference between transition 1 and 5, a larger $\delta$ should be introduced to achieve effective $\pi$ rotation. Supplementary Fig.~\ref{Sfig:Master simulation for h3} \textbf{d} shows a line cut at $\delta$=0.28 meV. 

We note that, although the simulation demonstrates the capability for coherent manipulation of higher energy orbital states, the number of allowable operations is limited by the rapid population relaxation, which are only $\sim$28~ps and 22~ps~\cite{Yan2023} for $\left|h_3\right>$ and $\left|h_4\right>$, respectively. A longer orbital relaxation time could be achieved by tuning the orbital-state spacings via controlling the QD growth conditions~\cite{Zibik2009,Pan2000}.

\begin{figure}[h]
\refstepcounter{fig}
	\includegraphics[width=1\linewidth]{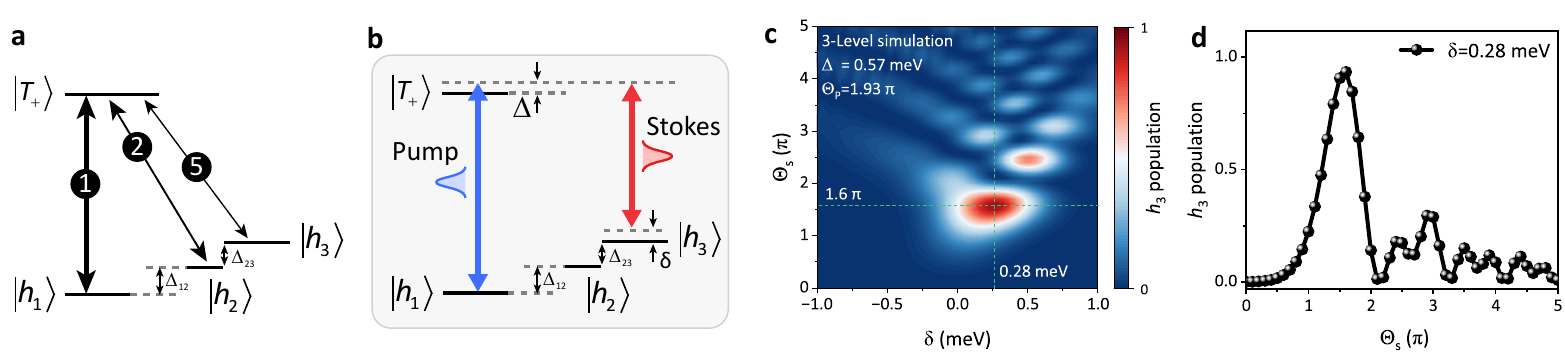}
	\caption{Maser equation simulation of SRT process between $\left|h_1\right>$ and $\left|h_3\right>$. (a) Energy diagram showing the involved transitions and (b) the corresponding pulses. (c) Simulated contour plot as functions of two-photon detuning ($\delta$) and pulse area ($\Theta_s$). (d) Line cut at $\delta$= 0.28 meV.}
\label{Sfig:Master simulation for h3}
\end{figure}

\newpage
\subsection{Modeling of pump-probe experiments with noise}
\label{section: with noise}
\begin{figure}[h]
\refstepcounter{fig}
	\includegraphics[width=1\linewidth]{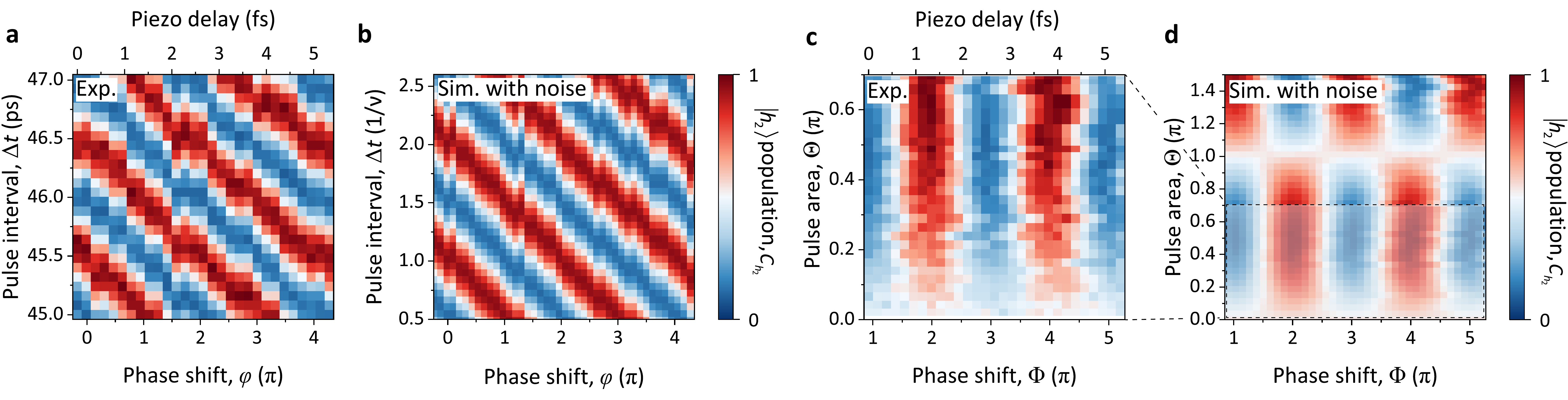}
	\caption{Lindblad master equation simulations of two Raman pulse pump-probe experiments with noise. Experimentally recorded (a, c) and simulated (b, d) $h_2$ population contour plots: (a, b) as functions of pulse interval and phase shift; (c, d) as functions of pulse area and phase shift.}

\label{Sfig:Master simulation with jitter}
\end{figure}

To better reproduce the experimental results, we account for the noise present in practical experiments. Given that the optical setup experiences vibrations during the experiment, we consider Gaussian-distributed fluctuations in the initial phase of the pulse and the phase scanning span, with an initial phase fluctuation full-width at half-maximum (FWHM) of 0.037$\pi$ and a phase scanning span fluctuation FWHM corresponding to 1.8\% of the phase values, respectively. Additionally, we consider pulse area fluctuations with a Gaussian distribution and an FWHM value corresponding to 0.54\% of the pulse area value. The pure dephasing and population relaxation processes are introduced using Lindblad dissipators. Consequently, the evolution of the system's density matrix is described by the Lindblad master equation:

\begin{equation}
i\hbar\frac{d\rho(\textit{t})}{d\textit{t}}=[H(t), \rho(t)]+i\hbar\frac{\gamma_1}{2} \mathcal{L}[A_1] \rho+i\hbar \frac{\gamma_2}{2} \mathcal{L}[A_2] \rho,
\end{equation}

where $\mathcal{L}[A] \rho=2 A \rho A^{\dagger}-A^{\dagger} A \rho-\rho A^{\dagger} A$ is the Lindblad superoperator. $A_1=\left|h_2\right>\left<h_2\right|$ and $A_2=\left|h_1\right>\left<h_2\right|$ are the dissipators representing pure dephasing and population relaxation, with rates $\gamma_1=\frac{1}{263}$~ps$^{-1}$ and $\gamma_2=\frac{1}{159}$~ps$^{-1}$, respectively. Supplementary Fig.~\ref{Sfig:Master simulation with jitter} presents the simulated contour plots as well as the corresponding experimental results for straightforward comparison. 

\newpage
\section{Evaluation of the Raman pulse fidelity}
\label{Ssection:Raman pulse fidelity}
To assess the fidelity of the Raman pulse with pulse area of $\pi$, we measure the emission intensity following a two-step coherent population transfer process  from $\left|h_1\right>$ to $\left|h_2\right>$. Firstly, we pump the population to $\left|T_+\right>$ (purple arrow in Supplementary Fig.~\ref{Sfig:Resonant CPT}a). Secondly, we stimulate the Auger emission (blue arrow) to $\left|h_2\right>$ with a fidelity of 95\%, as shown in Supplementary Fig.~\ref{Sfig:Resonant CPT}b. Finally, the readout CW laser is introduced to determine the intensity expected to be detected when $\left|h_2\right>$ population is $100\%$ (dashed line in Supplementary Fig.~\ref{Sfig:Resonant CPT}d).

\begin{figure}[h]
\includegraphics[width=0.55\textwidth]{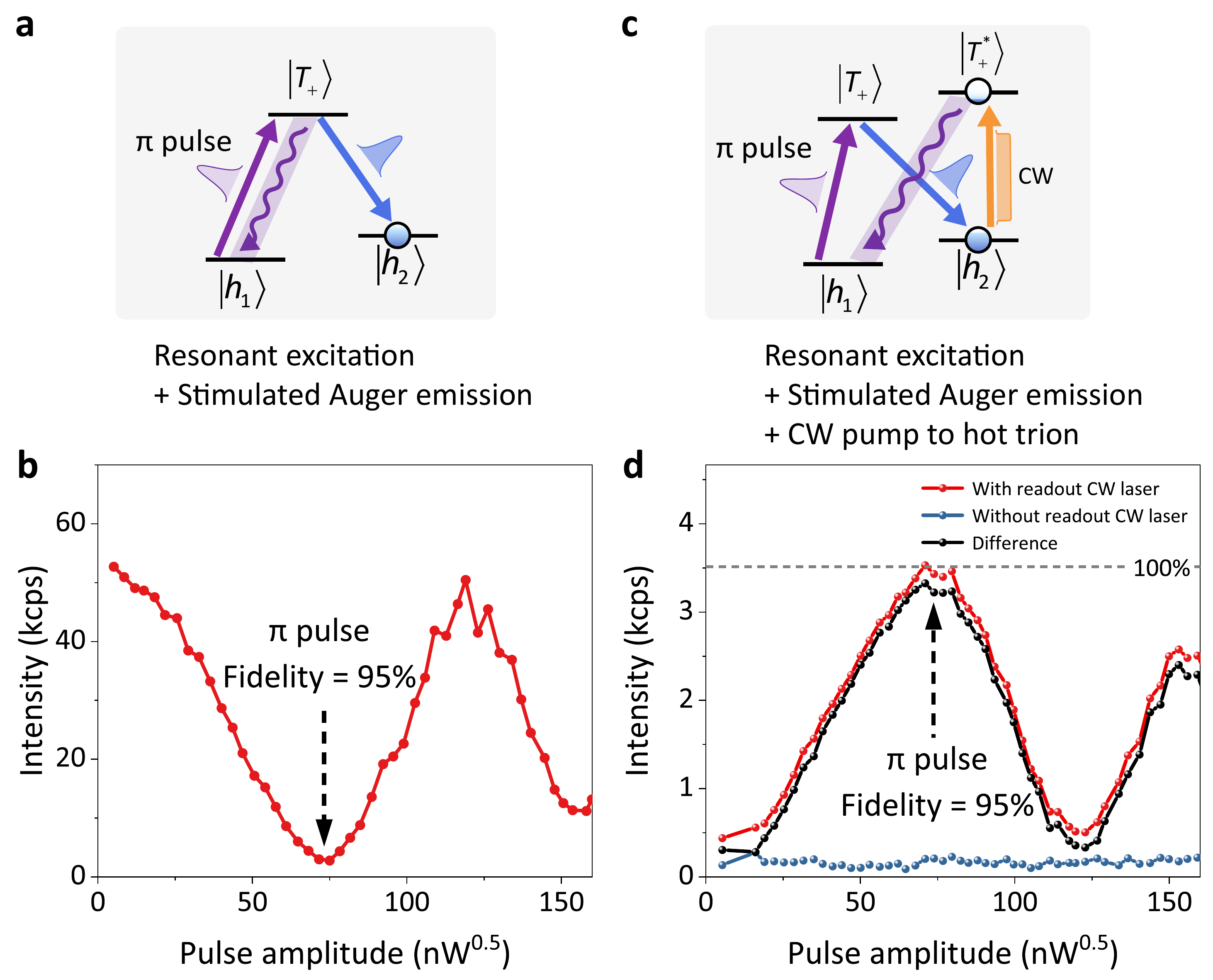}
\caption{\textbf{a}, Schematic of $\pi$ pulse resonant excitation of $\left|T_+\right>$ and then stimulating the Auger emission and detection the $\left|T_+\right>\rightarrow\left|\textit{h}_\text{1}\right>$ emission. \textbf{b}, Detected photon emission from $\left|T_+\right>\rightarrow\left|h_1\right>$ transition as a function of pulse amplitude (blue arrow in \textbf{c}). By comparing the intensity at stimulated pulse area $\Theta=\pi$ and $\Theta=0$, we obtained the $\left|h_2\right>$ preparation fidelity $F=95\%$. \textbf{c}, Same as \textbf{a} but introducing a readout CW laser and detecting the $\left|T_+^*\right>\rightarrow\left|h_1\right>$ emission. \textbf{d}, Detected photon emission from $\left|T_+^*\right>\rightarrow\left|h_1\right>$ transition as a function of pulse amplitude (blue arrow in \textbf{c}), from which we obtain the intensity expected to be detected when $\left|h_2\right>$ population is $100\%$ (dashed horizontal line). Red (blue) dots: detected intensity with (without) CW laser. Black dots: the difference in intensity with and without CW laser.}
\label{Sfig:Resonant CPT}
\end{figure}

To achieve maximum Raman pulse fidelity, we continuously increase the Stokes pulse amplitude for three different pump pulse amplitudes. As shown in Supplementary Fig.~\ref{Sfig:Rabi at diff pump power}, we measure the emission intensity as a function of Stokes pulse amplitude with (red) and without (blue) readout CW laser. The emission observed without the readout CW laser arises from the unintended excitation of $\left|T_+\right>$ and $\left|T_+^*\right>$ due to phonon-assisted excitation~\cite{Quilter2015a} and breakdown of the adiabatic elimination approximation~\cite{Press2008}. We take the difference (black) between measured intensity with and without CW laser as the real population of $\left|h_2\right>$ induced by the stimulated Raman process. 
By analyzing this data for different pump pulse amplitudes ($24.5~\rm{nW^{0.5}}$, $30.0~\rm{nW^{0.5}}$ and $34.6~\rm{nW^{0.5}}$), we achieve $\pi$ pulse fidelity of $F_{\pi}=84.0\%$, 87.3\% and 87.0\%, respectively. Additionally, the pulse amplitude ratios $E_{\rm{pump}}/E_{\rm{Stokes}}$ at the $\pi$ pulse condition vary from 3.2 to 5.1, which aligns well with the ratio of dipole moments measured in a separate experiment ($\mu_{1}/\mu_{2}=4.8$, Supplementary Fig.~\ref{Sfig:dipole moment}).

\begin{figure}[h]
\includegraphics[width=0.95\textwidth]{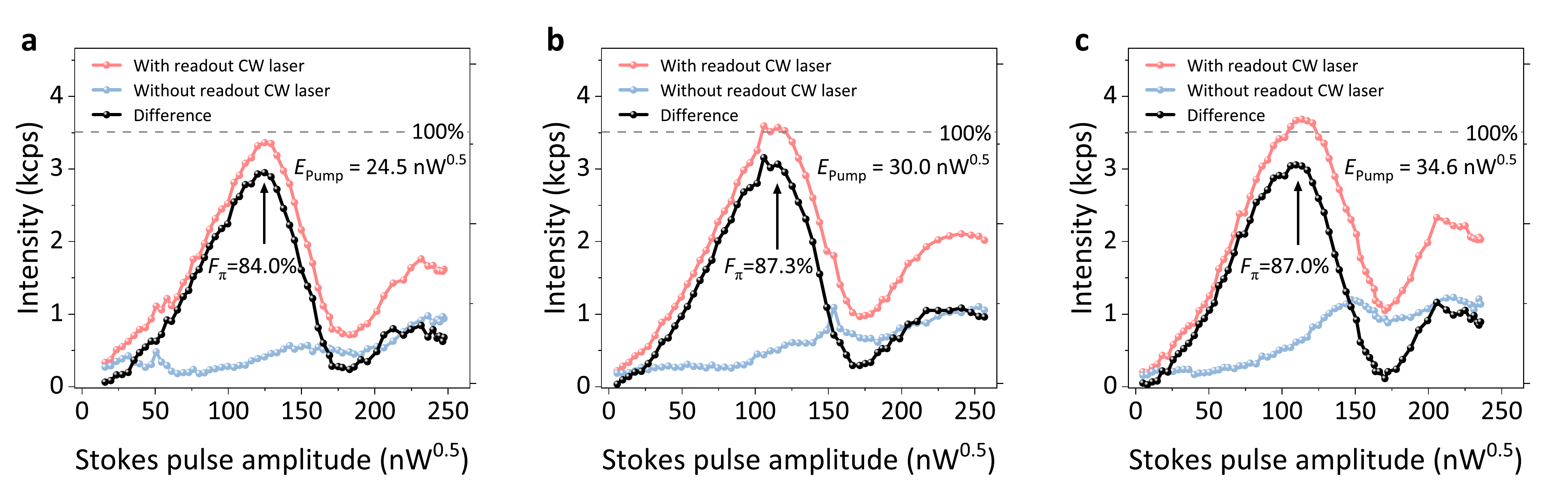}
\caption{Detected $\left|T_+^*\right>\rightarrow\left|\textit{h}_\text{1}\right>$ emission intensity as a function of Stokes pulse amplitude for pump pulse amplitude fixed at $24.5~\rm{nW^{0.5}}$ (\textbf{a}), $30.0~\rm{nW^{0.5}}$ (\textbf{b}), $34.6~\rm{nW^{0.5}}$ (\textbf{c}). Light red (blue): detected intensity with (without) readout CW laser, respectively. Black: the difference in intensity with and without CW laser.}
\label{Sfig:Rabi at diff pump power}
\end{figure}

\newpage
\section{Lifetime measurement of the orbital qubit}
\label{Section-lifetime}
We measure the lifetime, $T_1$, of the orbital qubit by a three-pulse pump-probe technique. The lifetime is given by $T_1=1/(\frac{1}{T_{1,h_1}}+\frac{1}{T_{1,h_2}})$, where $T_{1,h_1}$, $T_{1,h_2}$ are the lifetime of $\left|h_1\right>$ and $\left|h_2\right>$, respectively. As $\left|h_1\right>$ is expected to have a relatively long lifetime of a few microseconds~\cite{Yan2023}, we can reasonably assume that $T_1$ is approximately equal to $T_{1,h_2}$. As shown in Supplementary Fig.~\ref{Sfig:T1}, we first prepare a $\left|h_2\right>$ by a sequence of two $\pi$ pulses. Subsequently, a delayed readout pulse is employed to measure the population of $\left|h_2\right>$. As we vary the pulse interval, $\Delta t$, we obtain a 159-ps $\left|h_2\right>$ lifetime, corresponding $2T_1=318(4)$~ps.

\begin{figure}[h]
\includegraphics[width=0.6\textwidth]{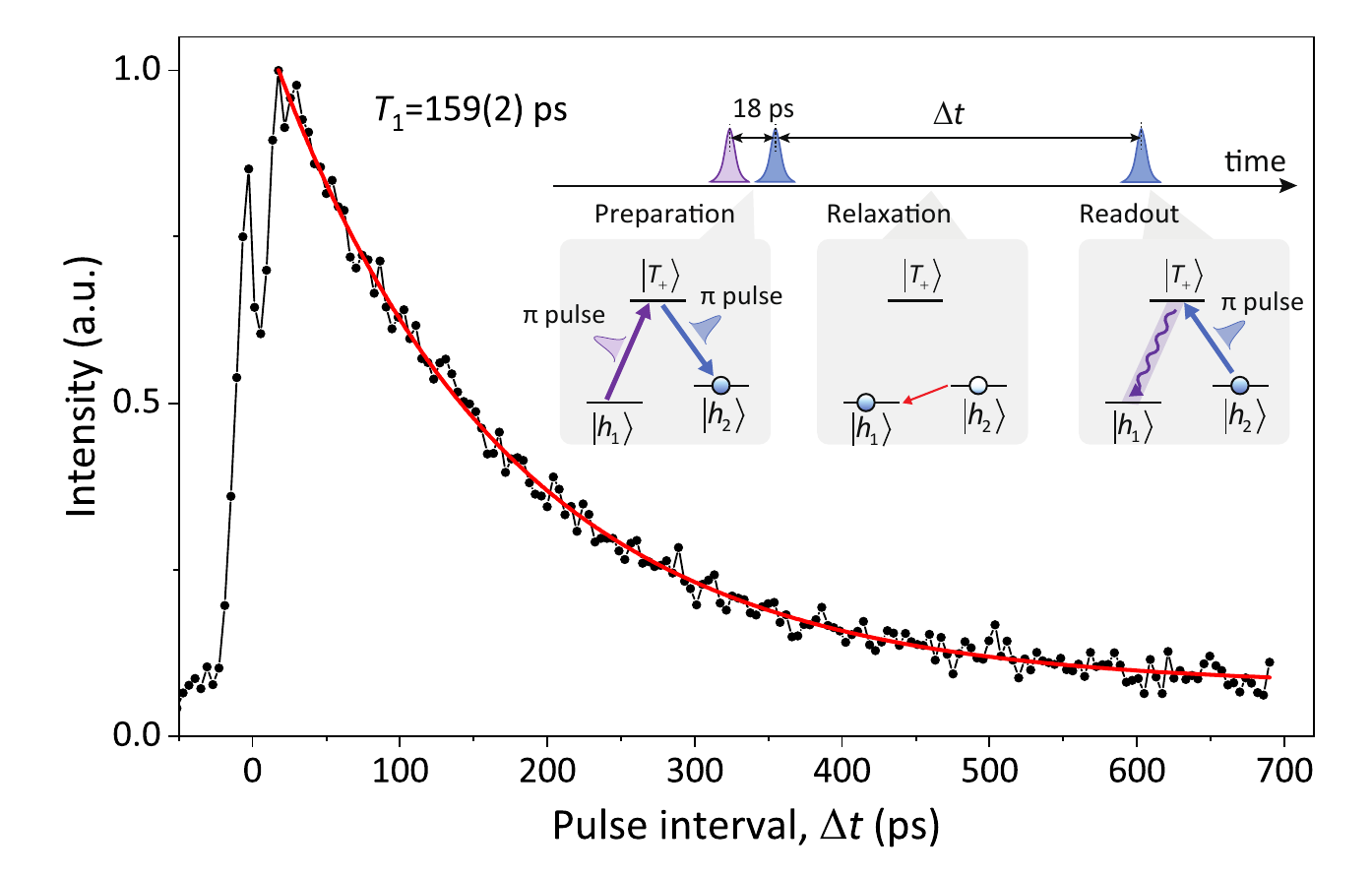}
\caption{Measurement of $\left|h_2\right>$ lifetime, $T_1$.}
\label{Sfig:T1}
\end{figure}

\newpage
\section{Experimental setup}
\label{sec:sample}
We present a schematic of the experimental setup in Supplementary Fig.~\ref{Sfig:setup}. The sample is located in a close-cycle cryostat (attocube) with a base temperature of 3.5 K. We use a cross-polarized microscope setup and notch filters (OptiGrate) to filter out Raman pulses and CW laser scattering. For CW readout, a tunable narrow-linewidth Ti:sapphire laser (M Squared) is used. For Raman pulses, we use two folded 4$f$ pulse shapers to pick out phase-locked pulses with picosecond duration and different colours operating as pump and Stokes pulses. The femtosecond pulse is generated from a tunable mode-locked Ti:sapphire laser (Coherent) with a pulse duration of 140~fs at a repetition rate of 80~MHz. The power of each pulse is independently controlled by a rotational ND filter (LBTEK). The pulse interval between the control and probe Raman pulses is introduced by a motorized optical delay line (Newport). The optical phase of the control Raman pulse is adjusted by a closed-loop piezo (CoreMorrow) with a resolution of 0.2~nm. Photons emitted from the QD are collected using a single-mode fiber and directed to a spectrometer for spectral characterization. The spectrometer (Princeton Instruments) has a 1800 lines/mm grating, a 750 mm focal length, and a spectral resolution of $\sim$30 ueV.
\begin{figure}[h]
\refstepcounter{fig}
	\includegraphics[width=1\linewidth]{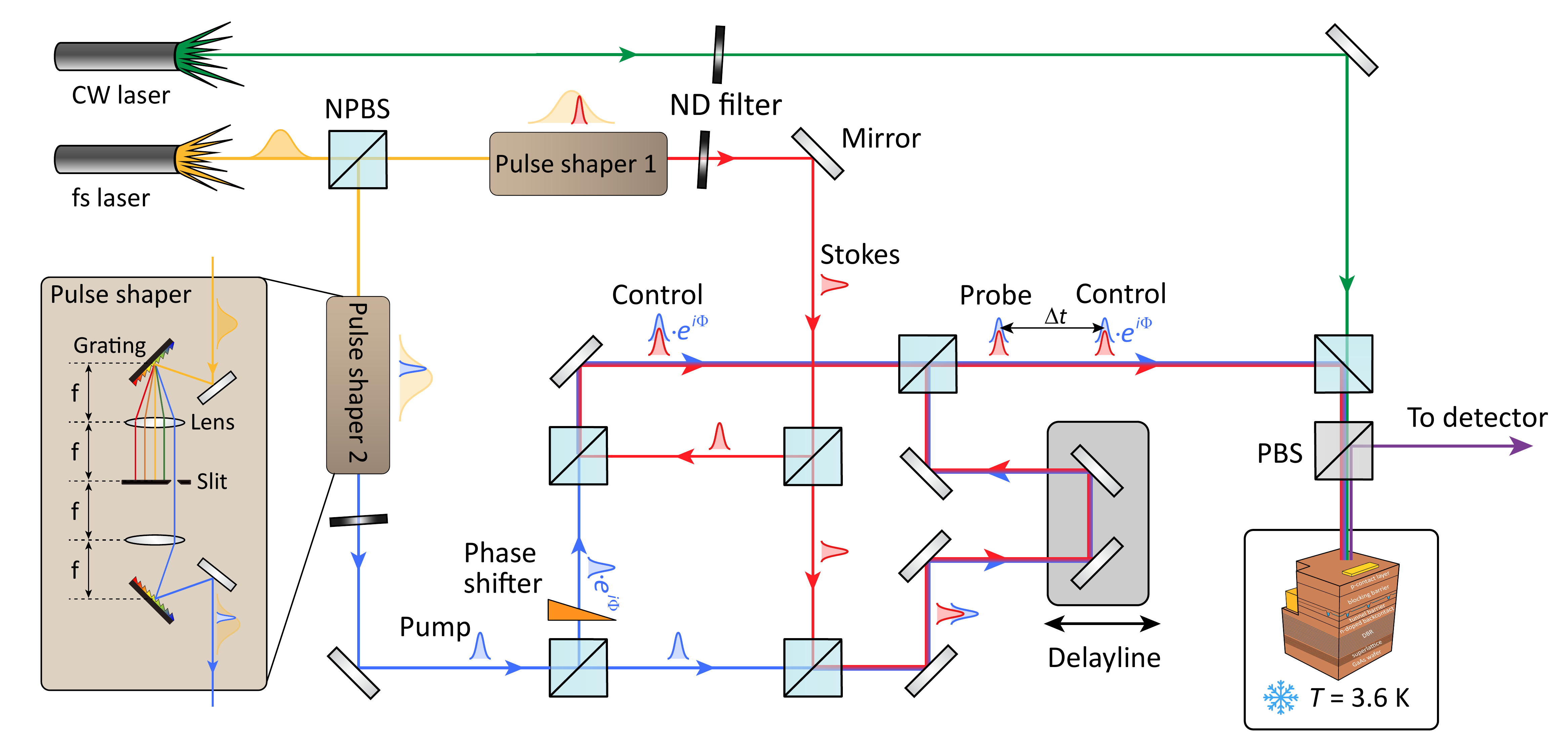}
	\caption{Schematic of the experimental setup. NPBS: non-polarizing beam splitter. PBS: polarizing beam splitter. ND filter: neutral density filter. Green: readout CW laser. Blue (red): pump (Stokes) pulse. Purple: signal.}
\label{Sfig:setup}
\end{figure}
\putbib

\end{bibunit}


\begin{thebibliography}{10}
\expandafter\ifx\csname url\endcsname\relax
  \def\url#1{\texttt{#1}}\fi
\expandafter\ifx\csname urlprefix\endcsname\relax\def\urlprefix{URL }\fi
\providecommand{\bibinfo}[2]{#2}
\providecommand{\eprint}[2][]{\url{#2}}

\bibitem{Ramakrishnan2023}
\bibinfo{author}{Wehner, S.}, \bibinfo{author}{Elkouss, D.} \& \bibinfo{author}{Hanson, R.}
\newblock \bibinfo{title}{{Quantum internet: A vision for the road ahead}}.
\newblock \emph{\bibinfo{journal}{Science}} \textbf{\bibinfo{volume}{362}}, \bibinfo{pages}{eaam9288} (\bibinfo{year}{2018}).

\bibitem{Lu2021b}
\bibinfo{author}{Lu, C.-Y.} \& \bibinfo{author}{Pan, J.-W.}
\newblock \bibinfo{title}{{Quantum-dot single-photon sources for the quantum internet}}.
\newblock \emph{\bibinfo{journal}{Nature Nanotechnology}} \textbf{\bibinfo{volume}{16}}, \bibinfo{pages}{1294--1296} (\bibinfo{year}{2021}).

\bibitem{Walther2005}
\bibinfo{author}{Walther, P.} \emph{et~al.}
\newblock \bibinfo{title}{{Experimental one-way quantum computing}}.
\newblock \emph{\bibinfo{journal}{Nature}} \textbf{\bibinfo{volume}{434}}, \bibinfo{pages}{169--176} (\bibinfo{year}{2005}).

\bibitem{Lindner2009}
\bibinfo{author}{Lindner, N.~H.} \& \bibinfo{author}{Rudolph, T.}
\newblock \bibinfo{title}{{Proposal for Pulsed On-Demand Sources of Photonic Cluster State Strings}}.
\newblock \emph{\bibinfo{journal}{Physical Review Letters}} \textbf{\bibinfo{volume}{103}}, \bibinfo{pages}{113602} (\bibinfo{year}{2009}).

\bibitem{Economou2010a}
\bibinfo{author}{Economou, S.~E.}, \bibinfo{author}{Lindner, N.} \& \bibinfo{author}{Rudolph, T.}
\newblock \bibinfo{title}{{Optically Generated 2-Dimensional Photonic Cluster State from Coupled Quantum Dots}}.
\newblock \emph{\bibinfo{journal}{Physical Review Letters}} \textbf{\bibinfo{volume}{105}}, \bibinfo{pages}{093601} (\bibinfo{year}{2010}).

\bibitem{Blatt2008}
\bibinfo{author}{Blatt, R.} \& \bibinfo{author}{Wineland, D.}
\newblock \bibinfo{title}{{Entangled states of trapped atomic ions}}.
\newblock \emph{\bibinfo{journal}{Nature}} \textbf{\bibinfo{volume}{453}}, \bibinfo{pages}{1008--1015} (\bibinfo{year}{2008}).

\bibitem{Thomas2022}
\bibinfo{author}{Thomas, P.}, \bibinfo{author}{Ruscio, L.}, \bibinfo{author}{Morin, O.} \& \bibinfo{author}{Rempe, G.}
\newblock \bibinfo{title}{{Efficient generation of entangled multiphoton graph states from a single atom}}.
\newblock \emph{\bibinfo{journal}{Nature}} \textbf{\bibinfo{volume}{608}}, \bibinfo{pages}{677--681} (\bibinfo{year}{2022}).

\bibitem{Yang2022}
\bibinfo{author}{Yang, C.-W.} \emph{et~al.}
\newblock \bibinfo{title}{{Sequential generation of multiphoton entanglement with a Rydberg superatom}}.
\newblock \emph{\bibinfo{journal}{Nature Photonics}} \textbf{\bibinfo{volume}{16}}, \bibinfo{pages}{658--661} (\bibinfo{year}{2022}).

\bibitem{Bernien2013}
\bibinfo{author}{Bernien, H.} \emph{et~al.}
\newblock \bibinfo{title}{{Heralded entanglement between solid-state qubits separated by three metres}}.
\newblock \emph{\bibinfo{journal}{Nature}} \textbf{\bibinfo{volume}{497}}, \bibinfo{pages}{86--90} (\bibinfo{year}{2013}).

\bibitem{Zhai2022}
\bibinfo{author}{Zhai, L.} \emph{et~al.}
\newblock \bibinfo{title}{{Quantum interference of identical photons from remote GaAs quantum dots}}.
\newblock \emph{\bibinfo{journal}{Nature Nanotechnology}} \textbf{\bibinfo{volume}{17}}, \bibinfo{pages}{829--833} (\bibinfo{year}{2022}).

\bibitem{Ding2016d}
\bibinfo{author}{Ding, X.} \emph{et~al.}
\newblock \bibinfo{title}{{On-Demand Single Photons with High Extraction Efficiency and Near-Unity Indistinguishability from a Resonantly Driven Quantum Dot in a Micropillar}}.
\newblock \emph{\bibinfo{journal}{Physical Review Letters}} \textbf{\bibinfo{volume}{116}}, \bibinfo{pages}{020401} (\bibinfo{year}{2016}).

\bibitem{Huber2017a}
\bibinfo{author}{Huber, D.} \emph{et~al.}
\newblock \bibinfo{title}{{Highly indistinguishable and strongly entangled photons from symmetric GaAs quantum dots}}.
\newblock \emph{\bibinfo{journal}{Nature Communications}} \textbf{\bibinfo{volume}{8}}, \bibinfo{pages}{15506} (\bibinfo{year}{2017}).

\bibitem{Somaschi2016}
\bibinfo{author}{Somaschi, N.} \emph{et~al.}
\newblock \bibinfo{title}{{Near-optimal single-photon sources in the solid state}}.
\newblock \emph{\bibinfo{journal}{Nature Photonics}} \textbf{\bibinfo{volume}{10}}, \bibinfo{pages}{340--345} (\bibinfo{year}{2016}).

\bibitem{Liu2018f}
\bibinfo{author}{Liu, F.} \emph{et~al.}
\newblock \bibinfo{title}{{High Purcell factor generation of indistinguishable on-chip single photons}}.
\newblock \emph{\bibinfo{journal}{Nature Nanotechnology}} \textbf{\bibinfo{volume}{13}}, \bibinfo{pages}{835--840} (\bibinfo{year}{2018}).

\bibitem{Liu2019e}
\bibinfo{author}{Liu, J.} \emph{et~al.}
\newblock \bibinfo{title}{{A solid-state source of strongly entangled photon pairs with high brightness and indistinguishability}}.
\newblock \emph{\bibinfo{journal}{Nature Nanotechnology}} \textbf{\bibinfo{volume}{14}}, \bibinfo{pages}{586--593} (\bibinfo{year}{2019}).

\bibitem{Tiranov2023}
\bibinfo{author}{Tiranov, A.} \emph{et~al.}
\newblock \bibinfo{title}{{Collective super- and subradiant dynamics between distant optical quantum emitters}}.
\newblock \emph{\bibinfo{journal}{Science}} \textbf{\bibinfo{volume}{379}}, \bibinfo{pages}{389--393} (\bibinfo{year}{2023}).

\bibitem{DeGreve2012a}
\bibinfo{author}{{De Greve}, K.} \emph{et~al.}
\newblock \bibinfo{title}{{Quantum-dot spin–photon entanglement via frequency downconversion to telecom wavelength}}.
\newblock \emph{\bibinfo{journal}{Nature}} \textbf{\bibinfo{volume}{491}}, \bibinfo{pages}{421--425} (\bibinfo{year}{2012}).

\bibitem{Gao2012}
\bibinfo{author}{Gao, W.~B.}, \bibinfo{author}{Fallahi, P.}, \bibinfo{author}{Togan, E.}, \bibinfo{author}{Miguel-Sanchez, J.} \& \bibinfo{author}{Imamoglu, A.}
\newblock \bibinfo{title}{{Observation of entanglement between a quantum dot spin and a single photon}}.
\newblock \emph{\bibinfo{journal}{Nature}} \textbf{\bibinfo{volume}{491}}, \bibinfo{pages}{426--430} (\bibinfo{year}{2012}).

\bibitem{Coste2023}
\bibinfo{author}{Coste, N.} \emph{et~al.}
\newblock \bibinfo{title}{{High-rate entanglement between a semiconductor spin and indistinguishable photons}}.
\newblock \emph{\bibinfo{journal}{Nature Photonics}} \textbf{\bibinfo{volume}{17}}, \bibinfo{pages}{582--587} (\bibinfo{year}{2023}).

\bibitem{Cogan2023}
\bibinfo{author}{Cogan, D.}, \bibinfo{author}{Su, Z.-E.}, \bibinfo{author}{Kenneth, O.} \& \bibinfo{author}{Gershoni, D.}
\newblock \bibinfo{title}{{Deterministic generation of indistinguishable photons in a cluster state}}.
\newblock \emph{\bibinfo{journal}{Nature Photonics}} \textbf{\bibinfo{volume}{17}}, \bibinfo{pages}{324--329} (\bibinfo{year}{2023}).

\bibitem{Appel2022}
\bibinfo{author}{Appel, M.~H.} \emph{et~al.}
\newblock \bibinfo{title}{{Entangling a Hole Spin with a Time-Bin Photon: A Waveguide Approach for Quantum Dot Sources of Multiphoton Entanglement}}.
\newblock \emph{\bibinfo{journal}{Physical Review Letters}} \textbf{\bibinfo{volume}{128}}, \bibinfo{pages}{233602} (\bibinfo{year}{2022}).

\bibitem{Delteil2016a}
\bibinfo{author}{Delteil, A.} \emph{et~al.}
\newblock \bibinfo{title}{{Generation of heralded entanglement between distant hole spins}}.
\newblock \emph{\bibinfo{journal}{Nature Physics}} \textbf{\bibinfo{volume}{12}}, \bibinfo{pages}{218--223} (\bibinfo{year}{2016}).

\bibitem{Yan2023}
\bibinfo{author}{Yan, J.-Y.} \emph{et~al.}
\newblock \bibinfo{title}{{Coherent control of a high-orbital hole in a semiconductor quantum dot}}.
\newblock \emph{\bibinfo{journal}{Nature Nanotechnology}} \textbf{\bibinfo{volume}{18}}, \bibinfo{pages}{1139--1146} (\bibinfo{year}{2023}).

\bibitem{Monroe1995}
\bibinfo{author}{Monroe, C.}, \bibinfo{author}{Meekhof, D.~M.}, \bibinfo{author}{King, B.~E.}, \bibinfo{author}{Itano, W.~M.} \& \bibinfo{author}{Wineland, D.~J.}
\newblock \bibinfo{title}{{Demonstration of a Fundamental Quantum Logic Gate}}.
\newblock \emph{\bibinfo{journal}{Physical Review Letters}} \textbf{\bibinfo{volume}{75}}, \bibinfo{pages}{4714--4717} (\bibinfo{year}{1995}).

\bibitem{Press2008}
\bibinfo{author}{Press, D.}, \bibinfo{author}{Ladd, T.~D.}, \bibinfo{author}{Zhang, B.} \& \bibinfo{author}{Yamamoto, Y.}
\newblock \bibinfo{title}{{Complete quantum control of a single quantum dot spin using ultrafast optical pulses}}.
\newblock \emph{\bibinfo{journal}{Nature}} \textbf{\bibinfo{volume}{456}}, \bibinfo{pages}{218--221} (\bibinfo{year}{2008}).

\bibitem{Press2010a}
\bibinfo{author}{Press, D.} \emph{et~al.}
\newblock \bibinfo{title}{{Ultrafast optical spin echo in a single quantum dot}}.
\newblock \emph{\bibinfo{journal}{Nature Photonics}} \textbf{\bibinfo{volume}{4}}, \bibinfo{pages}{367--370} (\bibinfo{year}{2010}).

\bibitem{Gao2015}
\bibinfo{author}{Gao, W.~B.}, \bibinfo{author}{Imamoglu, A.}, \bibinfo{author}{Bernien, H.} \& \bibinfo{author}{Hanson, R.}
\newblock \bibinfo{title}{{Coherent manipulation, measurement and entanglement of individual solid-state spins using optical fields}}.
\newblock \emph{\bibinfo{journal}{Nature Photonics}} \textbf{\bibinfo{volume}{9}}, \bibinfo{pages}{363--373} (\bibinfo{year}{2015}).

\bibitem{Spinnler2021}
\bibinfo{author}{Spinnler, C.} \emph{et~al.}
\newblock \bibinfo{title}{{Optically driving the radiative Auger transition}}.
\newblock \emph{\bibinfo{journal}{Nature Communications}} \textbf{\bibinfo{volume}{12}}, \bibinfo{pages}{6575} (\bibinfo{year}{2021}).

\bibitem{Lobl2020}
\bibinfo{author}{L{\"{o}}bl, M.~C.} \emph{et~al.}
\newblock \bibinfo{title}{{Radiative Auger process in the single-photon limit}}.
\newblock \emph{\bibinfo{journal}{Nature Nanotechnology}} \textbf{\bibinfo{volume}{15}}, \bibinfo{pages}{558--562} (\bibinfo{year}{2020}).

\bibitem{Gawarecki2023}
\bibinfo{author}{Gawarecki, K.} \emph{et~al.}
\newblock \bibinfo{title}{{Symmetry breaking via alloy disorder to explain radiative Auger transitions in self-assembled quantum dots}}.
\newblock \emph{\bibinfo{journal}{Physical Review B}} \textbf{\bibinfo{volume}{108}}, \bibinfo{pages}{235410} (\bibinfo{year}{2023}).

\bibitem{Bodey2019}
\bibinfo{author}{Bodey, J.~H.} \emph{et~al.}
\newblock \bibinfo{title}{{Optical spin locking of a solid-state qubit}}.
\newblock \emph{\bibinfo{journal}{npj Quantum Information}} \textbf{\bibinfo{volume}{5}}, \bibinfo{pages}{95} (\bibinfo{year}{2019}).

\bibitem{Berezovsky2008}
\bibinfo{author}{Berezovsky, J.}, \bibinfo{author}{Mikkelsen, M.~H.}, \bibinfo{author}{Stoltz, N.~G.}, \bibinfo{author}{Coldren, L.~A.} \& \bibinfo{author}{Awschalom, D.~D.}
\newblock \bibinfo{title}{{Picosecond Coherent Optical Manipulation of a Single Electron Spin in a Quantum Dot}}.
\newblock \emph{\bibinfo{journal}{Science}} \textbf{\bibinfo{volume}{320}}, \bibinfo{pages}{349--352} (\bibinfo{year}{2008}).

\bibitem{Buckley2010}
\bibinfo{author}{Buckley, B.~B.}, \bibinfo{author}{Fuchs, G.~D.}, \bibinfo{author}{Bassett, L.~C.} \& \bibinfo{author}{Awschalom, D.~D.}
\newblock \bibinfo{title}{{Spin-Light Coherence for Single-Spin Measurement and Control in Diamond}}.
\newblock \emph{\bibinfo{journal}{Science}} \textbf{\bibinfo{volume}{330}}, \bibinfo{pages}{1212--1215} (\bibinfo{year}{2010}).

\bibitem{Zhou2017}
\bibinfo{author}{Zhou, B.~B.} \emph{et~al.}
\newblock \bibinfo{title}{{Holonomic Quantum Control by Coherent Optical Excitation in Diamond}}.
\newblock \emph{\bibinfo{journal}{Physical Review Letters}} \textbf{\bibinfo{volume}{119}}, \bibinfo{pages}{140503} (\bibinfo{year}{2017}).

\bibitem{Sweeney2011a}
\bibinfo{author}{Sweeney, T.~M.}, \bibinfo{author}{Phelps, C.} \& \bibinfo{author}{Wang, H.}
\newblock \bibinfo{title}{{Quantum control of electron spins in the two-dimensional electron gas of a CdTe quantum well with a pair of Raman-resonant phase-locked laser pulses}}.
\newblock \emph{\bibinfo{journal}{Physical Review B}} \textbf{\bibinfo{volume}{84}}, \bibinfo{pages}{075321} (\bibinfo{year}{2011}).

\bibitem{Petta2004}
\bibinfo{author}{Petta, J.~R.}, \bibinfo{author}{Johnson, A.~C.}, \bibinfo{author}{Marcus, C.~M.}, \bibinfo{author}{Hanson, M.~P.} \& \bibinfo{author}{Gossard, A.~C.}
\newblock \bibinfo{title}{{Manipulation of a Single Charge in a Double Quantum Dot}}.
\newblock \emph{\bibinfo{journal}{Physical Review Letters}} \textbf{\bibinfo{volume}{93}}, \bibinfo{pages}{186802} (\bibinfo{year}{2004}).

\bibitem{Cao2013}
\bibinfo{author}{Cao, G.} \emph{et~al.}
\newblock \bibinfo{title}{{Ultrafast universal quantum control of a quantum-dot charge qubit using Landau–Zener–St{\"{u}}ckelberg interference}}.
\newblock \emph{\bibinfo{journal}{Nature Communications}} \textbf{\bibinfo{volume}{4}}, \bibinfo{pages}{1401} (\bibinfo{year}{2013}).

\bibitem{Descamps2023}
\bibinfo{author}{Descamps, T.} \emph{et~al.}
\newblock \bibinfo{title}{{Semiconductor Membranes for Electrostatic Exciton Trapping in Optically Addressable Quantum Transport Devices}}.
\newblock \emph{\bibinfo{journal}{Physical Review Applied}} \textbf{\bibinfo{volume}{19}}, \bibinfo{pages}{044095} (\bibinfo{year}{2023}).

\bibitem{Fujita2019}
\bibinfo{author}{Fujita, T.} \emph{et~al.}
\newblock \bibinfo{title}{{Angular momentum transfer from photon polarization to an electron spin in a gate-defined quantum dot}}.
\newblock \emph{\bibinfo{journal}{Nature Communications}} \textbf{\bibinfo{volume}{10}}, \bibinfo{pages}{2991} (\bibinfo{year}{2019}).

\bibitem{Babin2022}
\bibinfo{author}{Babin, H.-G.} \emph{et~al.}
\newblock \bibinfo{title}{{Full wafer property control of local droplet etched GaAs quantum dots}}.
\newblock \emph{\bibinfo{journal}{Journal of Crystal Growth}} \textbf{\bibinfo{volume}{591}}, \bibinfo{pages}{126713} (\bibinfo{year}{2022}).

\bibitem{Stemmann2008a}
\bibinfo{author}{Stemmann, A.}, \bibinfo{author}{Heyn, C.}, \bibinfo{author}{K{\"{o}}ppen, T.}, \bibinfo{author}{Kipp, T.} \& \bibinfo{author}{Hansen, W.}
\newblock \bibinfo{title}{{Local droplet etching of nanoholes and rings on GaAs and AlGaAs surfaces}}.
\newblock \emph{\bibinfo{journal}{Applied Physics Letters}} \textbf{\bibinfo{volume}{93}}, \bibinfo{pages}{123108} (\bibinfo{year}{2008}).

\bibitem{Zhai2020}
\bibinfo{author}{Zhai, L.} \emph{et~al.}
\newblock \bibinfo{title}{{Low-noise GaAs quantum dots for quantum photonics}}.
\newblock \emph{\bibinfo{journal}{Nature Communications}} \textbf{\bibinfo{volume}{11}}, \bibinfo{pages}{4745} (\bibinfo{year}{2020}).

\bibitem{Schimpf2021}
\bibinfo{author}{Schimpf, C.}, \bibinfo{author}{Manna, S.}, \bibinfo{author}{{Covre da Silva}, S.~F.}, \bibinfo{author}{Aigner, M.} \& \bibinfo{author}{Rastelli, A.}
\newblock \bibinfo{title}{{Entanglement-based quantum key distribution with a blinking-free quantum dot operated at a temperature up to 20 K}}.
\newblock \emph{\bibinfo{journal}{Advanced Photonics}} \textbf{\bibinfo{volume}{3}}, \bibinfo{pages}{065001} (\bibinfo{year}{2021}).

\bibitem{Warburton2000}
\bibinfo{author}{Warburton, R.~J.} \emph{et~al.}
\newblock \bibinfo{title}{{Optical emission from a charge-tunable quantum ring}}.
\newblock \emph{\bibinfo{journal}{Nature}} \textbf{\bibinfo{volume}{405}}, \bibinfo{pages}{926--929} (\bibinfo{year}{2000}).

\bibitem{SI}
\bibinfo{title}{{See Supplemental Material for further information about the experimental details and discussions of the modeling, which additionally includes Refs.~\cite{Nguyen2012a,Johansson2012a,Zibik2009,Pan2000,Quilter2015a}}} .

\bibitem{Tinkey2022}
\bibinfo{author}{Tinkey, H.~N.}, \bibinfo{author}{Clark, C.~R.}, \bibinfo{author}{Sawyer, B.~C.} \& \bibinfo{author}{Brown, K.~R.}
\newblock \bibinfo{title}{{Transport-Enabled Entangling Gate for Trapped Ions}}.
\newblock \emph{\bibinfo{journal}{Physical Review Letters}} \textbf{\bibinfo{volume}{128}}, \bibinfo{pages}{050502} (\bibinfo{year}{2022}).

\bibitem{Chathanathil2023}
\bibinfo{author}{Chathanathil, J.}, \bibinfo{author}{Ramaswamy, A.}, \bibinfo{author}{Malinovsky, V.~S.}, \bibinfo{author}{Budker, D.} \& \bibinfo{author}{Malinovskaya, S.~A.}
\newblock \bibinfo{title}{{Chirped fractional stimulated Raman adiabatic passage}}.
\newblock \emph{\bibinfo{journal}{Physical Review A}} \textbf{\bibinfo{volume}{108}}, \bibinfo{pages}{043710} (\bibinfo{year}{2023}).

\bibitem{Greilich2011}
\bibinfo{author}{Greilich, A.}, \bibinfo{author}{Carter, S.~G.}, \bibinfo{author}{Kim, D.}, \bibinfo{author}{Bracker, A.~S.} \& \bibinfo{author}{Gammon, D.}
\newblock \bibinfo{title}{{Optical control of one and two hole spins in interacting quantum dots}}.
\newblock \emph{\bibinfo{journal}{Nature Photonics}} \textbf{\bibinfo{volume}{5}}, \bibinfo{pages}{702--708} (\bibinfo{year}{2011}).

\bibitem{Godden2012}
\bibinfo{author}{Godden, T.~M.} \emph{et~al.}
\newblock \bibinfo{title}{{Coherent Optical Control of the Spin of a Single Hole in an InAs/GaAs Quantum Dot}}.
\newblock \emph{\bibinfo{journal}{Physical Review Letters}} \textbf{\bibinfo{volume}{108}}, \bibinfo{pages}{017402} (\bibinfo{year}{2012}).

\bibitem{Ringbauer2021}
\bibinfo{author}{Ringbauer, M.} \emph{et~al.}
\newblock \bibinfo{title}{{A universal qudit quantum processor with trapped ions}}.
\newblock \emph{\bibinfo{journal}{Nature Physics}} \textbf{\bibinfo{volume}{18}}, \bibinfo{pages}{1053--1057} (\bibinfo{year}{2022}).

\bibitem{Chi2022}
\bibinfo{author}{Chi, Y.} \emph{et~al.}
\newblock \bibinfo{title}{{A programmable qudit-based quantum processor}}.
\newblock \emph{\bibinfo{journal}{Nature Communications}} \textbf{\bibinfo{volume}{13}}, \bibinfo{pages}{1166} (\bibinfo{year}{2022}).

\bibitem{Tiurev2021}
\bibinfo{author}{Tiurev, K.} \emph{et~al.}
\newblock \bibinfo{title}{{Fidelity of time-bin-entangled multiphoton states from a quantum emitter}}.
\newblock \emph{\bibinfo{journal}{Physical Review A}} \textbf{\bibinfo{volume}{104}}, \bibinfo{pages}{052604} (\bibinfo{year}{2021}).

\bibitem{Wu2019}
\bibinfo{author}{Wu, Y.} \emph{et~al.}
\newblock \bibinfo{title}{{Observation of parity-time symmetry breaking in a single-spin system}}.
\newblock \emph{\bibinfo{journal}{Science}} \textbf{\bibinfo{volume}{364}}, \bibinfo{pages}{878--880} (\bibinfo{year}{2019}).

\bibitem{Antolinez2019}
\bibinfo{author}{Antolinez, F.~V.}, \bibinfo{author}{Rabouw, F.~T.}, \bibinfo{author}{Rossinelli, A.~A.}, \bibinfo{author}{Cui, J.} \& \bibinfo{author}{Norris, D.~J.}
\newblock \bibinfo{title}{{Observation of Electron Shakeup in CdSe/CdS Core/Shell Nanoplatelets}}.
\newblock \emph{\bibinfo{journal}{Nano Letters}} \textbf{\bibinfo{volume}{19}}, \bibinfo{pages}{8495--8502} (\bibinfo{year}{2019}).

\bibitem{Llusar2020}
\bibinfo{author}{Llusar, J.} \& \bibinfo{author}{Climente, J.~I.}
\newblock \bibinfo{title}{{Nature and Control of Shakeup Processes in Colloidal Nanoplatelets}}.
\newblock \emph{\bibinfo{journal}{ACS Photonics}} \textbf{\bibinfo{volume}{7}}, \bibinfo{pages}{3086--3095} (\bibinfo{year}{2020}).

\bibitem{Dean1967}
\bibinfo{author}{Dean, P.~J.}, \bibinfo{author}{Cuthbert, J.~D.}, \bibinfo{author}{Thomas, D.~G.} \& \bibinfo{author}{Lynch, R.~T.}
\newblock \bibinfo{title}{{Two-Electron Transitions in the Luminescence of Excitons Bound to Neutral Donors in Gallium Phosphide}}.
\newblock \emph{\bibinfo{journal}{Physical Review Letters}} \textbf{\bibinfo{volume}{18}}, \bibinfo{pages}{122--124} (\bibinfo{year}{1967}).

\bibitem{Bryja2016}
\bibinfo{author}{Bryja, L.} \emph{et~al.}
\newblock \bibinfo{title}{{Thermal dissociation of free and acceptor-bound positive trions from magnetophotoluminescence studies of high quality GaAs/Al$_x$Ga$_{1-x}$As quantum wells}}.
\newblock \emph{\bibinfo{journal}{Physical Review B}} \textbf{\bibinfo{volume}{93}}, \bibinfo{pages}{165303} (\bibinfo{year}{2016}).

\bibitem{Seyler2019}
\bibinfo{author}{Seyler, K.~L.} \emph{et~al.}
\newblock \bibinfo{title}{{Signatures of moir{\'{e}}-trapped valley excitons in MoSe$_2$/WSe$_2$ heterobilayers}}.
\newblock \emph{\bibinfo{journal}{Nature}} \textbf{\bibinfo{volume}{567}}, \bibinfo{pages}{66--70} (\bibinfo{year}{2019}).

\bibitem{Alexeev2019}
\bibinfo{author}{Alexeev, E.~M.} \emph{et~al.}
\newblock \bibinfo{title}{{Resonantly hybridized excitons in moir{\'{e}} superlattices in van der Waals heterostructures}}.
\newblock \emph{\bibinfo{journal}{Nature}} \textbf{\bibinfo{volume}{567}}, \bibinfo{pages}{81--86} (\bibinfo{year}{2019}).

\bibitem{Baek2020}
\bibinfo{author}{Baek, H.} \emph{et~al.}
\newblock \bibinfo{title}{{Highly energy-tunable quantum light from moir{\'{e}}-trapped excitons}}.
\newblock \emph{\bibinfo{journal}{Science Advances}} \textbf{\bibinfo{volume}{6}}, \bibinfo{pages}{eaba8526} (\bibinfo{year}{2020}).

\bibitem{Binder2019}
\bibinfo{author}{Binder, J.} \emph{et~al.}
\newblock \bibinfo{title}{{Upconverted electroluminescence via Auger scattering of interlayer excitons in van der Waals heterostructures}}.
\newblock \emph{\bibinfo{journal}{Nature Communications}} \textbf{\bibinfo{volume}{10}}, \bibinfo{pages}{2335} (\bibinfo{year}{2019}).

\bibitem{Nguyen2012a}
\bibinfo{author}{Nguyen, H.~S.} \emph{et~al.}
\newblock \bibinfo{title}{{Optically Gated Resonant Emission of Single Quantum Dots}}.
\newblock \emph{\bibinfo{journal}{Physical Review Letters}} \textbf{\bibinfo{volume}{108}}, \bibinfo{pages}{057401} (\bibinfo{year}{2012}).

\bibitem{Johansson2012a}
\bibinfo{author}{Johansson, J.}, \bibinfo{author}{Nation, P.} \& \bibinfo{author}{Nori, F.}
\newblock \bibinfo{title}{{QuTiP: An open-source Python framework for the dynamics of open quantum systems}}.
\newblock \emph{\bibinfo{journal}{Computer Physics Communications}} \textbf{\bibinfo{volume}{183}}, \bibinfo{pages}{1760--1772} (\bibinfo{year}{2012}).

\bibitem{Zibik2009}
\bibinfo{author}{Zibik, E.~A.} \emph{et~al.}
\newblock \bibinfo{title}{{Long lifetimes of quantum-dot intersublevel transitions in the terahertz range}}.
\newblock \emph{\bibinfo{journal}{Nature Materials}} \textbf{\bibinfo{volume}{8}}, \bibinfo{pages}{803--807} (\bibinfo{year}{2009}).

\bibitem{Pan2000}
\bibinfo{author}{Pan, D.}, \bibinfo{author}{Towe, E.}, \bibinfo{author}{Kennerly, S.} \& \bibinfo{author}{Kong, M.-Y.}
\newblock \bibinfo{title}{{Tuning of conduction intersublevel absorption wavelengths in (In, Ga)As/GaAs quantum-dot nanostructures}}.
\newblock \emph{\bibinfo{journal}{Applied Physics Letters}} \textbf{\bibinfo{volume}{76}}, \bibinfo{pages}{3537--3539} (\bibinfo{year}{2000}).

\bibitem{Quilter2015a}
\bibinfo{author}{Quilter, J.~H.} \emph{et~al.}
\newblock \bibinfo{title}{{Phonon-Assisted Population Inversion of a Single InGaAs/GaAs Quantum Dot by Pulsed Laser Excitation}}.
\newblock \emph{\bibinfo{journal}{Physical Review Letters}} \textbf{\bibinfo{volume}{114}}, \bibinfo{pages}{137401} (\bibinfo{year}{2015}).

\end{thebibliography}


\begin{thebibliography}{1}
\expandafter\ifx\csname url\endcsname\relax
  \def\url#1{\texttt{#1}}\fi
\expandafter\ifx\csname urlprefix\endcsname\relax\def\urlprefix{URL }\fi
\providecommand{\bibinfo}[2]{#2}
\providecommand{\eprint}[2][]{\url{#2}}

\bibitem{Nguyen2012a}
\bibinfo{author}{Nguyen, H.~S.} \emph{et~al.}
\newblock \bibinfo{title}{{Optically Gated Resonant Emission of Single Quantum Dots}}.
\newblock \emph{\bibinfo{journal}{Physical Review Letters}} \textbf{\bibinfo{volume}{108}}, \bibinfo{pages}{057401} (\bibinfo{year}{2012}).

\bibitem{Johansson2012a}
\bibinfo{author}{Johansson, J.}, \bibinfo{author}{Nation, P.} \& \bibinfo{author}{Nori, F.}
\newblock \bibinfo{title}{{QuTiP: An open-source Python framework for the dynamics of open quantum systems}}.
\newblock \emph{\bibinfo{journal}{Computer Physics Communications}} \textbf{\bibinfo{volume}{183}}, \bibinfo{pages}{1760--1772} (\bibinfo{year}{2012}).

\bibitem{Yan2023}
\bibinfo{author}{Yan, J.-Y.} \emph{et~al.}
\newblock \bibinfo{title}{{Coherent control of a high-orbital hole in a semiconductor quantum dot}}.
\newblock \emph{\bibinfo{journal}{Nature Nanotechnology}} \textbf{\bibinfo{volume}{18}}, \bibinfo{pages}{1139--1146} (\bibinfo{year}{2023}).

\bibitem{Zibik2009}
\bibinfo{author}{Zibik, E.~A.} \emph{et~al.}
\newblock \bibinfo{title}{{Long lifetimes of quantum-dot intersublevel transitions in the terahertz range}}.
\newblock \emph{\bibinfo{journal}{Nature Materials}} \textbf{\bibinfo{volume}{8}}, \bibinfo{pages}{803--807} (\bibinfo{year}{2009}).

\bibitem{Pan2000}
\bibinfo{author}{Pan, D.}, \bibinfo{author}{Towe, E.}, \bibinfo{author}{Kennerly, S.} \& \bibinfo{author}{Kong, M.-Y.}
\newblock \bibinfo{title}{{Tuning of conduction intersublevel absorption wavelengths in (In, Ga)As/GaAs quantum-dot nanostructures}}.
\newblock \emph{\bibinfo{journal}{Applied Physics Letters}} \textbf{\bibinfo{volume}{76}}, \bibinfo{pages}{3537--3539} (\bibinfo{year}{2000}).

\bibitem{Quilter2015a}
\bibinfo{author}{Quilter, J.~H.} \emph{et~al.}
\newblock \bibinfo{title}{{Phonon-Assisted Population Inversion of a Single InGaAs/GaAs Quantum Dot by Pulsed Laser Excitation}}.
\newblock \emph{\bibinfo{journal}{Physical Review Letters}} \textbf{\bibinfo{volume}{114}}, \bibinfo{pages}{137401} (\bibinfo{year}{2015}).

\bibitem{Press2008}
\bibinfo{author}{Press, D.}, \bibinfo{author}{Ladd, T.~D.}, \bibinfo{author}{Zhang, B.} \& \bibinfo{author}{Yamamoto, Y.}
\newblock \bibinfo{title}{{Complete quantum control of a single quantum dot spin using ultrafast optical pulses}}.
\newblock \emph{\bibinfo{journal}{Nature}} \textbf{\bibinfo{volume}{456}}, \bibinfo{pages}{218--221} (\bibinfo{year}{2008}).

\end{thebibliography}
\end{document}